%% file: sobolev-final.tex
\documentclass[10pt]{article}

\usepackage{epsfig,cancel,amsmath,amsthm,amssymb}
\usepackage{color}

\usepackage{lscape}
\usepackage{wrapfig}
\usepackage{color}
\usepackage{graphicx,framed}
\usepackage[colorlinks=true, pdfstartview=FitV, linkcolor=black, citecolor=blue, urlcolor=blue]{hyperref}
\definecolor{shadecolor}{rgb}{0.95, 0.95, 0.86}

\def\red#1{\textcolor[rgb]{0.9, 0, 0}{#1} }

\def\blue#1{\textcolor[rgb]{0,0,1}{#1}}

\def \gg{{\mathfrak g}}
\def\replace#1#2{\red{\cancel{{\hbox{\scriptsize  {$#1$}} }}}{\blue {#2}}}

\def \wt{\widetilde}

\def\ra{\rightarrow}

\def\Id{ \mathrm{Id}}

\newcommand{\ga}{\gamma} 
\newcommand{\G}{\Gamma} 
\renewcommand{\O}{\Omega} 
\renewcommand{\k}{\varkappa} 
\renewcommand{\d}{\delta}
 
\newcommand{\Si}{\Sigma} 
\newcommand{\e}{\epsilon}
\renewcommand{\o}{\omega}  
\newcommand{\g}{\gamma} 
\newcommand{\la}{\lambda}

\newcommand{\pa}{\partial}       
       
\newcommand{\coi}{C_0^{\infty}}

\newcommand{\br}{{\mathbb R}}

\newcommand{\CH}{{\mathcal H}}

\newtheorem{theorem}{Theorem}[section]
\newtheorem{example}[theorem]{Example}
\newtheorem{exercise}[theorem]{Exercise}

\newtheorem{lemma}[theorem]{Lemma}
\newtheorem{remark}[theorem]{Remark}
\newtheorem{problem}[theorem]{Riemann-Hilbert Problem}

\newtheorem{proposition}[theorem]{Proposition} 
\newtheorem{corollary}[theorem]{Corollary} 
\newtheorem{definition}[theorem]{Definition}
\def\ri{\right}
\def\ds{\displaystyle}

\def\res{\mathop{\mathrm {res}}\limits_}

\def\bth{\begin{theorem}}
\def\et{\end{theorem}}
\def\bc{\begin{corollary}}
\def\ec{\end{corollary}}
\def\bx{\begin{example}\small}
\def\ex{\end{example}}
\def\bxr{\begin{exercise}\small}
\def\exr{\end{exercise}}
\def\bl{\begin{lemma}}
\def\el{\end{lemma}}
\def\bd{\begin{definition}}
\def\ed{\end{definition}}
\def\bp{\begin{proposition}}
\def\ep{\end{proposition}}

\def\br{\begin{remark}}
\def\er{\end{remark}}

\def\be{\begin{equation}}
\def\ee{\end{equation}}
\def\ov {\overline}

\def\&{\hspace{-15pt}&}
\def\bea{\begin{eqnarray}}
\def\eea{\end{eqnarray}}
\def\beas{\begin{eqnarray*}}
\def\eeas{\end{eqnarray*}}
\def\bi{\begin{itemize}}
\def\ei{\end{itemize}}

\def \pa{\partial}
\def\C{{\mathbb C}}
\def\L{\mathcal L}
\def\R{{\mathbb R}}
\def\N{{\mathbb N}}
\def\wh{\widehat}

\def\Z{{\mathbb Z}}
\def\u{\mathfrak u}

\def\l{\lambda}
\def\m{\mu}

\def\1{{\bf 1}}

\def\s{ {\sigma}} 
\def\t{ {\tau}} 
\def\tf{ {\tilde f}} 
\def\tk{ {\tilde \k}}
\def\tl{ {\tilde \l}}
 
\def\Th{ {\Theta}} 

\def\z{\zeta}

\def\hf{\frac{1}{2}}
\def\le{\left}
\def\ds{\displaystyle}

\newcommand{\Iscr}{\mathcal I}
\newcommand{\Jscr}{\mathcal J}

\newcommand{\Rscr}{\mathcal R}
\numberwithin{equation}{section}

\begin{document}
\baselineskip 18pt plus 1pt minus 1pt

\begin{center}
\begin{large}

\textbf{\sc On Sobolev instability of the interior problem of tomography}
\end{large}
\bigskip

M. Bertola$^\dagger$\footnote{The work was supported  in part by the Natural Sciences and Engineering Research Council of Canada. }
A. Katsevich$^\star$\footnote{The work  was supported in part by NSF grant DMS-1211164.}
 and 
 A. Tovbis $^\star$\footnote{The work  was supported in part by NSF grant DMS-1211164.}
 \bigskip
 
\begin{small}
$^{\dagger}$ {\em Centre de recherches math\'ematiques,
Universit\'e de Montr\'eal\\ C.~P.~6128, succ. centre ville, Montr\'eal,
Qu\'ebec, Canada H3C 3J7} and \\
 {\em  Department of Mathematics and
Statistics, Concordia University\\ 1455 de Maisonneuve W., Montr\'eal, Qu\'ebec,
Canada H3G 1M8} \\
\smallskip
$^{\star}$ {\em  University of Central Florida
	Department of Mathematics\\
	4000 Central Florida Blvd.
	P.O. Box 161364
	Orlando, FL 32816-1364
} \\
\end{small}
\end{center}
{\it E-mail:} bertola@mathstat.concordia.ca, Alexander.Katsevich@ucf.edu, Alexander.Tovbis@ucf.edu

\begin{abstract}
In this paper we continue investigation of the interior problem of tomography that was started in  \cite{BKT2}. As is known, solving the interior problem {with prior data specified on a finite collection of intervals $I_i$} is equivalent to analytic continuation of a function from $I_i$ to an open set ${\bf J}$. In the paper we prove that this analytic continuation can be obtained with the help of a simple explicit formula, which involves summation of a series. 
Our second result is that the operator of analytic continuation is not stable for any pair of Sobolev spaces  regardless of how close the set ${\bf J}$ is to $I_i$. Our main tool is the singular value decomposition of the operator $\CH^{-1}_e$ that arises when the interior problem is reduced to a problem of inverting the Hilbert transform from incomplete data. 
The asymptotics of the singular values and singular functions of $\CH^{-1}_e$, the latter being valid uniformly
on compact subsets {of the interior of $I_i$},  was  obtained in \cite{BKT2}. {Using these asymptotics we can accurately measure the degree of ill-posedness of the analytic continuation as a function of the target interval ${\bf J}$.} Our {last} result is the  convergence of the asymptotic approximation of the singular functions {in the $L^2(I_i)$ sense}. 
\end{abstract} 

\baselineskip 12pt


\section{Introduction}\label{math-intro}

Suppose one is interested in imaging a small region of interest (ROI) inside an object using tomography. In order to acquire a complete data set that enables stable reconstruction, one needs to send multiple x-rays through the object from many different directions. In particular, the x-rays that do not pass through the ROI are required as well. 
The interior problem of tomography arises when only the x-rays through the ROI are measured. In this case the tomographic data are incomplete, and image reconstruction becomes a challenging problem. In what follows, image reconstruction from x-ray data taylored to an ROI will be called the interior problem, and the corresponding data will be called interior data. Practical importance of the interior problem is clear, since tayloring the x-ray exposure to an ROI results in a reduced x-ray dose to the patient in medical applications of tomography. See \cite{wy-13} for a nice review of the state of the art in interior tomography.

One of the most powerfull tools for investigating the interior problem from the theoretical point of view is the Gelfand-Graev formula, which relates the tomographic data of an object with its one-dimensional Hilbert transform along lines \cite{gegr-91}. With the help of this formula, the interior problem of tomography can be reduced to the problem of inverting the Hilbert transform from incomplete data. 

Pick any line $L$ through the object. We regard $L$ as the $x$-axis. Fix some $2g+2$, $g\in\N$, distinct points $a_i$ on $L$: $a_i<a_{i+1}$, $i=1,2,\dots,2g+1$. Points $a_1$ and $a_{2g+2}$ mark the boundaries of the support of $f$ along $L$. Points $a_2$ and $a_{2g+1}$ mark the boundaries of the ROI along $L$. Consider the Finite Hilbert Transform (FHT)
\begin{equation}\label{def-hilb}
(\CH f)(x):= \frac1\pi \int_{a_1}^{a_{2g+2}} \frac{f|_L(y)}{y-x}dy,\ f|_L\in L^2([a_1,a_{2g+2}]).
\end{equation}
Here $f|_L$ is the restriction of $f$ to $L$, and $\CH f$ is the one-dimensional Hilbert transform of $f|_L$. Throughout the paper the line $L$ is always the same, so with some abuse of notation we write $f$ instead of $f|_L$. In the case of interior tomographic data, the Gelfand-Graev formula allows computation of $\CH f$ only on $[a_2,a_{2g+1}]$, but not on all $[a_1,a_{2g+2}]$. Thus the interior problem of tomography is reduced to finding $f$ inside the ROI, i.e. on $[a_2,a_{2g+1}]$, by solving the equation
\begin{equation}\label{hilb-dataintro}
(\CH f)(x)=\varphi(x),\ x\in [a_2,a_{2g+1}].
\end{equation}

Consider the operator $\CH:\, L_2([a_1,a_{2g+2}])\to L_2([a_2,a_{2g+1}])$. Unique recovery of $f$ on $[a_2,a_{2g+1}]$ is impossible since $\CH$ has a non-trivial kernel (see \cite{kt12} for its complete description). Therefore, to achieve unique recovery the data $\varphi$ should be augmented by some additional information. One type of information that guarantees uniqueness is the knowledge of $f$ on some {interval or}
intervals inside $[a_2,a_{2g+1}]$. This is the so-called interior problem with prior knowledge (\cite{yyw-07b, kcnd, cndk-08, wy-13})  {that will be considered below}. Let us assume that $f$ is known on the intervals 
\begin{equation}\label{int-int}
I_i:=[a_3,a_4]\cup [a_5,a_6]\cup\dots\cup [a_{2g-1},a_{2g}],
\end{equation}
which we call ``interior" (inside the ROI). Denote by $I_e:=[a_1,a_2]\cup [a_{2g+1},a_{2g+2}]$ the remaining ``exterior'' intervals (they are outside the ROI). Applying the FHT inversion formula (see e.g. \cite{oe91}), we get
\begin{equation}\label{hilb-inv}
\begin{split}
f(y)&=-\frac{w(y)}{\pi}\left(\int_{a_1}^{a_2}+\int_{a_{2g+1}}^{a_{2g+2}}\right) \frac{\varphi(x)}{w(x)(x-y)}dx-\frac{w(y)}{\pi}\int_{a_2}^{a_{2g+1}}\frac{\varphi(x)}{w(x)(x-y)}dx,\\
{\rm where}~~~w(x):&=\sqrt{(a_{2g+2}-x)(x-a_1)}~~~{\rm and}~~~\varphi(x)=(\CH f)(x),~ x\in [a_1,a_{2g+2}].
\end{split}
\end{equation}
The left side of (\ref{hilb-inv}) is known on $I_i$. The last integral on the right is known everywhere. Combining these known quantities we get an integral equation:
\begin{equation}\label{int-eq}
(\CH^{-1}_e\varphi)(y):=-\frac{w(y)}\pi \int_{I_e} \frac{\varphi(x)}{w(x)(x-y)}dx=\psi(y),\ y\in I_i,
\end{equation}
where 
\be\label{psi}
\psi(y)= f(y)+\frac{w(y)}{\pi}\int_{a_2}^{a_{2g+1}}\frac{\varphi(x)}{w(x)(x-y)}dx,\ y\in I_i
\ee
is a known function. 

The main problem we study in this paper is the stability of finding $f$ from the data. Several approaches to finding $f$ on $[a_2,a_{2g+1}]$ are possible. The first one consists of two steps. In step 1 we solve equation (\ref{int-eq}) for $\varphi(x)$ on  $I_e$. In step 2 we substitute the computed $\varphi(x)$ into (\ref{hilb-inv}) and recover $f(y)$ on $[a_2,a_{2g+1}]$. It is clear that solving (\ref{int-eq}), i.e. inverting $\CH^{-1}_e$, is the most unstable step. Consider the operator $\CH^{-1}_e$ in (\ref{int-eq}) as a map between two weighted $L^2$-spaces:
\begin{equation}\label{map}
\CH^{-1}_e:\ L^2(I_e,1/w)\to L^2(I_i,1/w).
\end{equation}
Its adjoint is the Hilbert transform:
\begin{equation}\label{hilb-adj}
(\CH_i\psi)(x):= \frac1\pi \int_{I_i} \frac{\psi(y)}{y-x}dy,\ x\in I_e.
\end{equation}
In \cite{BKT2} the authors studied the singular value decomposition (SVD) for the operator  $\CH^{-1}_e$.
Namely, we were interested in the singular values  $2\l=2\l_n>0$, $n\in\N$, and the corresponding left and right singular functions $f=f_n,~h=h_n$,  satisfying 
\begin{equation}\label{svd-def}
\begin{split}
(\CH^{-1}_e) h(y)=-\frac{w(y)}{\pi}&\int_{I_e} \frac{h(x)}{w(x)(x-y)}dx={2}\la f(y),\ y\in I_i,\\
(\CH_i f)(x)=\frac1\pi &\int_{I_i} \frac{f(y)}{y-x}dy={2}\la h(x),\ x\in I_e.
\end{split}
\end{equation}
See \eqref{svd-hat}--\eqref{straight} and Theorem~\ref{theo-whK}, which show that the SVD is well-defined. 
It is well known that the rate at which $\l_n$'s approach zero is related with the ill-posedness of inverting $\CH^{-1}_e$.
Because of the symmetry $(\l,f,h)\Leftrightarrow (-\l,-f,h)$ of \eqref{svd-def}, we are interested
only in positive $\l_n$. The main result of the paper \cite{BKT2} is the large $n$ asymptotics of $\l_n$, $f_n$ and $h_n$.

Let us introduce  a $g\times g$ matrix $\mathbb A$ by 
\be\label{matrixA}
(\mathbb A)_{kj}=2\int_{a_{2k}}^{a_{2k+1}}\frac{z^{j-1} dz}{R(z)}, ~~~k=1,\dots,g-1,~~~~{\rm and}~~~
(\mathbb A)_{gj}=2\int_{a_{1}}^{a_{2g+2}}\frac{z^{j-1} dz}{R_+(z)},~~~~j=1,\dots,g,
\ee
where $R(z)  = \prod_{j=1}^{2g+2}(z-a_j)^\hf$ is an analytic function on  $\C \setminus (I_e \cup I_i)$  behaving as $z^{g+1}$ at infinity, and
define
\be\label{tau11intro}
\t_{11}=-2\sum_{j=1}^g (\mathbb A^{-1})_{j1}\int_{I_e}\frac{z^{j-1} dz}{R_+(z)}.
\ee
Here and throughout the paper the subscripts $\pm$ routinely denote limiting values of  functions (vectors, matrices) from the left/right
side of  corresponding oriented arcs. In particular,
$R_+$ means the limiting value of $R$ on $I= I_e\cup I_i$ from $\Im z>0$. We also want to note that, according to
the well-known  Riemann's Theorem on periods of holomorphic differentials (\cite{FarkasKra},
$\tau_{11}$ is a purely imaginary number with positive imaginary part.
Then the asymptotics of $\l_n$ is given by (\cite{BKT2})
\be
\l_n  = {\rm e}^{-\frac {  i \pi}{\tau_{11}}n + \mathcal O(1)}\ ,\ \ n\to \infty. \label{in1}
\ee
 The asymptotics of the singular functions from \cite{BKT2}
 is described in Section \ref{sec-review} of this paper. An alternative approach to the analysis of SVD for the Hilbert transform with incomplete data is developed in \cite{kat10c, kat_11, kt12, aak13}.

The very rapid decay of singular values in \eqref{in1} indicates that finding $\varphi$ from $\psi$ is very unstable. This, however, does not imply that finding $f$ on $[a_2,a_{2g+1}]$ is unstable, since $f$ is computed by applying a smoothing operator to $\varphi$. The second approach to finding $f$ is based on the observation that the function $\psi$ defined by \eqref{psi} is analytic in $\mathbb C\setminus I_e$ (cf. \eqref{int-eq}). Hence, analytically continuing $\psi$ from $I_i$ to $(a_2,a_{2g+1})$, we can find $f$ using \eqref{psi} with $y\in (a_2,a_{2g+1})$. Note that any method that gives $f$ on $(a_2,a_{2g+1})$ is equivalend to analytic continuation of $\psi$ in view of \eqref{psi}. {\it Thus, analytic continuation of $\psi$ is at the heart of any method for solving the interior problem of tomography {with prior knowledge}.}

In this paper we obtain two results regarding the analytic continuation of $\psi$. We show that this analytic continuation can be obtained with the help of a simple explicit formula, which involves summation of a series, {see Corollary \ref{analcont-series}}. We prove that the series is absolutely convergent if $\psi$ is in the range of $\CH^{-1}_e$. We also analyze stability of this analytic continuation. Intuitively, it is clear that the farther away from $I_i$ we continue $\psi$ the less stable the procedure becomes. 
Our second result is that the operator of analytic continuation is not stable for any pair of Sobolev spaces: $H^{s_1}(I_i)\to H^{-s_2}(J)$, where $J$ is any open set containing $I_i$. In other words, the procedure is unstable no matter how close to $I_i$ we perform the continuation. This is an 
{interesting} result, because earlier related results indicated that finding $f$ might be stable \cite{dnck, kcnd}.

The paper is organized as follows. Since the derivation of our main results strongly depends on the results in \cite{BKT2}, the latter are briefly reviewed in Section~\ref{sec-review}. The analytic continuation of $\psi$ and its instability in the Sobolev spaces are established in Section~\ref{sec-sobol}. {Loosely speaking, this result shows that no matter how many derivatives are required of $\psi$, the continuation is not stable. The availability of asymptotics of singular values and singular functions allows us to accurately estimate the degree of instability of the continuation. In Section~\ref{sec-sobol} we introduce a Hilbert space $\mathcal A$ of functions defined on $I_i$ with the help of an exponentially growing weight. We show how fast this weight must grow in order to ensure that the analytic continuation from $I_i$ to an open set ${\bf J}$ be a continuous map from $\mathcal A\to L^2({\bf J})$. Thus, this rate of growth measures the degree of ill-posedness of the analytic continuation as a function of the target interval ${\bf J}$.}

In \cite{BKT2} it is shown that the asymptotic approximations to the exact singular functions $f_n$ are valid uniformly on compact subsets of the interior of $I_i$ as $n\to\infty$. In Section~\ref{sec-L2-conv} we show that these approximations are also valid in the $L^2(I_i)$ sense as well. This is the third result obtained in this paper. We do not consider the other set of singular functions that are defined on $I_e$, since they are not needed for the analytic continuation of $\psi$. The main idea of the approach in \cite{BKT2} is to reduce the SVD problem \eqref{svd-def} to a matrix Riemann-Hilbert problem (RHP), which, in turn, is asymptotically reduced to a simpler RHP. That simpler (model) RHP has an explicit solution, which can be expressed in terms of the Riemann Theta function. 
A brief review of the reduction to the model RHP  and 
certain related results from \cite{BKT2} are contained in Appendix~\ref{sec-ideas}. Some technical lemmas 
related to the approximation of singular functions on 
$[a_1,a_{2g+2}]\setminus I$ and on $I_i$
that are needed in Sections~\ref{sec-sobol} and \ref{sec-L2-conv} 
are proven in Appendix~\ref{proofstechn}.

\section{Brief review of main results of \cite{BKT2}}\label{sec-review}

This section contains a brief review of major results of  \cite{BKT2}. 
For convenience, most of the statements below are 
provided with direct references (in square brackets) to the corresponding  results of \cite{BKT2}.

The SVD system \eqref{svd-def} can be represented as
\bea\label{svd-hat}
(H^{-1}_e\wh h)(y)
&\&:=\frac{\sqrt{w(y)}}{2\pi i }\int_{I_e} \frac{\wh h(x)}{\sqrt{w(x)}(x-y)}dx
 = \la \wh f(y),\ y\in I_i,
\nonumber \\
(H_i\wh f)(x)&\&:=\frac1{2\pi i} \frac 1{\sqrt{w(x)}} \int_{I_i} \frac{\wh f(y)\sqrt{w(y)}}{(y-x)}dy=
\la \wh h(x),\ x\in I_e,
\label{svd-def2}
\eea
where  $\wh h   = \frac {h}{\sqrt{w}}\in L^2(I_e)$, $\wh f = \frac { i f}{\sqrt{w}}\in L^2(I_i)$, and the operators $H^{-1}_e$, $H_i$ act on the corresponding unweighted $L^2$ spaces. It can be checked directly  that the triple $(\l,\wh f,\wh h)$ satisfies the system \eqref{svd-def2} if and only if 
$\l,\psi$ is the eigenvalue/eigenvector of the integral operator  $(\hat K \phi)(z) =\int_I K(z,x)\phi(x)dx$ from $L^2(I)$ to $L^2(I)$, 
where
\be\label{svd-K}
K(z,x)=
\frac{w^{\frac 1 2}(x) w^{-\frac 1 2}(z)  \chi_e(z)\chi_i(x)  + w^{\frac 1 2}(z) w^{-\frac 1 2}(x)  
\chi_i(z)\chi_e(x)}{2i\pi (x-z)},~~ \psi =\wh f(z)  \chi_i(z)+\wh h(z)\chi_e(z).
\ee
(Here and henceforth $ \chi_i(z), \chi_e(z)$ denote the  characteristic (indicator) functions of the sets $I_i, I_e$, respectively.)
Thus, the SVD problem for the system \eqref{svd-def2} is reduced to the spectral problem for the integral operator $\wh K:L^2(I) \to L^2(I)$. 
It follows directly from \eqref{svd-K} that
\be
\wh K\big|_{L^2(I_i)}  = H_i\ ,\ \ \ \ \wh K\big|_{L^2(I_e)}  = H_e^{-1}.\label{straight}
\ee
\bth\label{theo-whK}[Thm.3.1 and  Cor.3.8]
$\wh K$ is a self-adjoint and a Hilbert--Schmidt operator. 
Moreover, all the eigenvalues of $\wh K$ are simple.
\et

According to Theorem \ref{theo-whK}, the eigenvalues of $\wh K$ are real
with the only possible point of accumulation $\l=0$. Since the singular values of \eqref{svd-def2} are positive (note the symmetry  
$(\l,\wh f,\wh h)\mapsto (-\l,-\wh f, \wh h)$ in \eqref{svd-def2}),  we are interested
only in the positive eigenvalues $\l_n$, $n\in\N$,  of $\wh K$, where we order $\l_0>\l_1>\dots >0$.

Let $\wh L$ denote the restrictions of $\wh K^2$ to the interval $I_i$. Then, according
to \eqref{svd-K}, $\wh L = H_e^{-1} H_i:L^2(I_i) \to L^2(I_i)$ is an integral operator
with eigenvalues $\l_n^2$ and eigenfunctions $\wh f_n$,  $n\in\N$. 
It is interesting to note 
(Lemma 3.6 in \cite{BKT2}) that $\wh L$ is a  strictly totally positive operator.
Then the simplicity of the eigenvalues  $\l^2_n$  of $\wh L$ and, thus, of  $\l_n$  of $\wh K$ in Theorem \ref{theo-whK}, follows from properties of 
 strictly totally positive integral operators (see \cite{Pinkus-Rev}). Another consequence of this property of $\wh L$
is that the singular function $\wh f_n$ has exactly $n$ sign changes on $I_i$, $n=0,1,2,\dots$.

An important object of the  spectral theory is the resolvent operator $\wh R$ of $\wh K$, defined by  
\be\label{whR}
(\Id + \wh R)(\Id -\frac 1 \l \wh K) = \Id.
\ee
The resolvent operator $\wh R$ is an integral operator with the kernel of the form 
\be
\label{resolvent}
R(z,x;\lambda) =   
\frac{
\vec g^t(x) \Gamma^{-1}(x;\lambda) \Gamma(z;\lambda)\vec f(z)} {2i\pi \l  (z-x)}, ~{\rm where}~
\vec f(z):= \le[
\begin{matrix}
{ \frac{i \chi_e(z)}{\sqrt{ w(z)}}} \\
\sqrt{ w(z)} \chi_i(z)
 \end{matrix}\ri], \ 
 \vec g(x):= \le[
 \begin{matrix}
-i\sqrt{ w(x)} \chi_i(x)\\
\frac{\chi_e(x)} {\sqrt{ w(x)}} 
\end{matrix}
\ri],
\ee
where $\vec g^t$ denotes the transposition of $\vec g$ and the matrix $\G(z;\l)$ satisfies the following 
Riemann-Hilbert Problem (RHP) \ref{RHPGamma}.

\begin{problem}
\label{RHPGamma}
Find a $2\times 2$ matrix-function $\G=\G(z;\la)$, $\l\in\C\setminus\{0\}$, which is  analytic in 
$\overline{\C}\setminus I$, where $I=I_i\cup I_e$, admits non-tangential boundary values   from the upper/lower half-planes that belong to $L^2_{loc}$ in the interior points of $I$, and satisfies 
\bea
\label{rhpGam}
\G_+(z;\l)&=\G_-(z;\l) \left[\begin{matrix} 1 & 0 \\ \frac{iw}{\la} & 1 \end{matrix}\right], \ \ z\in  I_i;\qquad 
\G_+(z;\l)=\G_-(z;\l) \left[\begin{matrix} 1 & -\frac{i}{\la w} \\ 0 & 1 \end{matrix}\right],\ \ z\in  I_e,
\\
\label{assGam}
&\G(z;\l)=\1+O(z^{-1})~~~~{\rm as} ~~z\ra\infty, \\
\label{endpcond-out}
&\G(z;\l)=\le[\mathcal O(1), \mathcal O((z-a_j)^{-\hf})\ri],~~z\ra a_j,~~j=1,2g+2,\\
\label{endpcond-out-inn}
&\G(z;\l)=\le[\mathcal O(1), \mathcal O(\ln(z-a_j))\ri],~~~~z\ra a_j,~~j=2,2g+1,\\
\label{endpcond-inn}
&\G(z;\l)=\le[\mathcal O(\ln (z-a_j)),\mathcal O(1)\ri], ~~~~~ z\ra a_j,~~j=3, \dots, 2g.
\eea
Here the endpoint  behavior of $\Gamma$ is described column-wise. We will frequently omit the dependence on $\l$ from notation and write simply $\G(z)$ for  convenience.
\end{problem}

The latest fact links the resolvent operator $\wh R$ for $\wh K$ with the RHP for  the matrix $\G$ from \eqref{resolvent}.

\bth\label{theo-Gam}[Thm.3.17 and Prop.3.12] 
The RHP \ref{RHPGamma} has a solution $\G(z;\l)$, where  $\l\in \C\setminus\{0\}$, if and only if $\l$ is not an eigenvalue 
of $\wh K$. Moreover, for any fixed  $\l\in \C\setminus\{0\}$ the   RHP \ref{RHPGamma} has at most one solution. 
\et

Connection of our spectral problem with the RHP \ref{RHPGamma} 
is remarkable, as  the RHP \ref{RHPGamma} is a much more convenient object for rigorous asymptotic analysis (in small $\l$) than
the spectral problem for $\wh K$. The eigenfunctions of $\wh K$ corresponding to a fixed eigenvalue $\l_n$ are given
by two proportional expressions
\be\label{round-phi_n}
\phi_{n,j}(z)=\frac{\chi_e(z)}{\sqrt{w(z)}} \res{\l=\l_n} \Gamma_{j1}(z;\l)  \frac {1}{\l} +
{i} \sqrt{w(z)}\chi_i(z)\res{\l=\l_n} \Gamma_{j2}(z;\l) \frac {1}{\l},~~~j=1,2,
\ee
in terms of the entries of the matrix $\G(z,\l)$,
where for every $n\in\N$ at least one  of $\phi_{n,j}$ is not identical zero on $I$. 


Once the connection between the spectral problem for $\wh K$ and the RHP \ref{RHPGamma} is established, we 
use the nonlinear steepest descent method of Deift and Zhou to construct an explicit 
leading order approximation of $\G(z,\l)$
as $\l\ra 0^+$  in terms of the Riemann Theta functions. Of course, this approximation will not be valid at the eigenvalues
$\l_n$ of $\wh K$, as, according to Theorem \ref{theo-Gam}, 
$\G(z,\l_n)$ does not exists. However, using the explicit form of the approximate solution, we can find
the values $\tl_n$ for which this approximate solution has singularities. The obtained values $\tl_n$ will be referred
to as ``approximate eigenvalues''. It tuns out that, indeed,  $\tl_n$ approximate the corresponding $\l_n$ with the
accuracy 
\be\label{acc_sing_val}
| \k_n -  \tk_n|=O( \tk_n^{-\frac 12}),
\ee 
where 
$\k_n=-\ln \l_n$ and  $\tk_n=-\ln \tl_n$ (it will be shown that $\tk_n=O(n)$ as $n\ra\infty$).

Let us now consider the asymptotics of singular functions.
According to \eqref{svd-K}, the approximation of normalized singular functions  can be obtained by replacing rows of 
the matrix
$\Gamma_{jk}(z;\l)$, $j,k\in\{1,2\}$, in \eqref{round-phi_n}
 by the corresponding rows of the approximate solution to the RHP \ref{RHPGamma}. 
To present the  approximation formula for singular functions,
we need to introduce some notations and a few notions from the theory of compact Riemann surfaces. They will also be helpful for a geometrical interpretation of $\tk_n$.

\begin{figure}
\begin{center}
\resizebox{0.9\textwidth}{!}{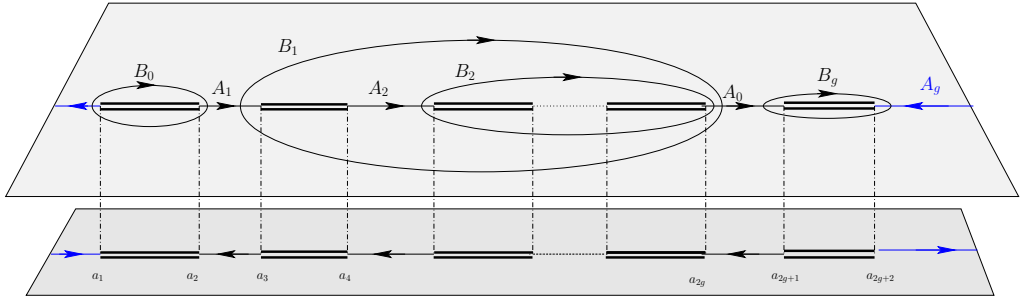}
\end{center}
\vspace{-10pt}
\caption{ Riemann surface  $\mathcal R$ with the choice of $A$ and $B$ cycles.}
\vspace{-12pt}
\label{homology}
\end{figure}

The {\bf Riemann Theta function} associated with a symmetric matrix $\tau$ with strictly positive imaginary part  
(that guarantees convergence) is the function of the vector argument $\vec z\in\C^g$ given by
\be\label{Theta}
\Theta(\vec z,\tau):= \sum_{\vec n\in \Z^g} \exp\bigg(i\pi   \vec n^t \cdot \tau \cdot \vec n +2i\pi \vec n^t \vec z\bigg).
\ee
Often the dependence on $\tau$ is  omitted from the notation. We will consider the matrix $\tau$  given by 
\be
\tau = [\tau_{ij}]= \le[
\oint_{B_i}\!\!\! \omega_j d\z
\ri]_{i,j=1,g},
\label{taumatrix}
\ee
where 
\be
\label{1stkind}
\vec \omega^t(z)= \le[
\begin{matrix}
\omega_1(z),\dots,
\omega_g(z)
\end{matrix}
\ri] = \frac { \le[\begin{array}{c}
1,\dots, z^{g-1}
\end{array}
\ri]}{R(z)}\mathbb A^{-1}, \ \ \ 
\ee
matrix $\mathbb A$ is defined by \eqref{matrixA}, and the loops (cycles) $B_i$, $i=1,\dots,g$ are shown in Figure \ref{homology}. 

\bth[Riemann \cite{FarkasKra}]
\label{Riemann1}
The matrix $\tau$ is {\bf symmetric} and its imaginary part is strictly positive definite.
\et

Matrix $\tau$ is an important object in the theory of compact Riemann surfaces. Indeed, consider the hyperelliptic Riemann surface  $\mathcal R$,
defined by the segments $[a_{2k-1},a_{2k}]$, $k=1,2,\dots,g+1$, that form $I$, with canonical $A$ and $B$ cycles shown in Figure \ref{homology}.
Then $\vec \omega(z)dz$ is known as the vector of normalized holomorphic differentials on $\mathcal R$ and $\tau$ is called 
the {normalized matrix of $B$-periods} of $\mathcal R$. Note that $[\mathbb A]_{ji}= 
\oint_{A_j} \frac {\z^{i-1} d \z}{R(\z)}$, and $\t_{11}$ in \eqref{tau11intro} is the $(1,1)$ entry of the matrix $\tau$. 

\br\label{rem-imag_tau}
It follows from \eqref{taumatrix}, \eqref{1stkind} and \eqref{matrixA} that the entries of the matrix $\t$ are purely imaginary.
\er

\bp
\label{thetaproperties}
For any $\lambda, \mu \in \Z^g$, the Theta function has the following properties:
\bea
&\& \Theta(\vec z,\tau)  = \Theta(-\vec z,\tau);\\
&\& \Theta(\vec z + \mu + \tau\lambda, \tau) = \exp \bigg( -2i\pi \lambda^t\vec z - i\pi  \lambda^t \tau \lambda \bigg)
\Theta(\vec z,\tau).\label{thetaperiods}
\eea
\ep

According to \eqref{Theta} and Proposition \ref{thetaproperties}, the Theta function is an even function
of $g$ complex variables, periodic on the lattice $\Z^g$ and  quasi-periodic  on the lattice $\t\Z^g$.
A hypersurface  $(\Th)\subset\C^g$, defined by $\Theta(\vec z,\tau)=0$, is called a theta divisor. 
This is a hypersurface of complex codimension one or real codimension two.
According to
Proposition \ref{thetaproperties}, the theta divisor $(\Th)$ is periodic in  $\Z^g$ and  $\t\Z^g$.

Let 
\be
 \label{wukappa} W= W(\k)=\frac {\k}{i\pi} \tau_1 + 2\mathfrak u(\infty)  + \frac {{\bf e}_1}{2},\ \ 
 W_0  = \frac {\tau_1}2 - \frac { {\bf e}_1 + {\bf e}_g}2,
  \ee
where $\t_1$ is the first column of matrix $\t$,
\be
\label{Abelmap}
\mathfrak u(z) = \int_{a_1}^z \vec \omega(\z)d\z  ,\ \ \ \ z\in \C \setminus [a_1,\infty),
\ee
is known as the Abel map on  $\mathcal R$, and ${\bf e}_k$  denotes the $k$th vector of the standard basis in $\C^g$.
Then  $\tk_n=-\ln \tl_n$ are defined by the condition
\be\label{kappa_n_cond}
 \Theta\le ( W(\k) - W_0\ri  )=0.
\ee
Geometrically, this condition  determines the points of intersection of the line $ W(\k)-W_0\subset \C^g$ with the theta divisor.
Let us consider this question in  a little more details. Direct calculations show that all the terms of $W(\k)$ in
\eqref{wukappa} are real, provided that $\k\in\R$. Thus, the line $\{W(\k):\k\in\R\}\subset \R^g\subset\R^{2g}$, if we identify
$\C^g$ with $\R^{2g}$. So, the line
 $ W(\k)-W_0$, $\k\in\R$, is a subset of the shifted hyperplane $\Pi=W_0+\R^g$. Let $(\Th)_R:=(\Th)\cap\Pi$.

\bl\label{lem-Theta_R}[Lem.7.5] 
Each connected component of  $(\Th)_R$ is a smooth  $g-1$  (real) dimensional hypersurface in $\Pi$.
\el

\begin{figure}[t]
\includegraphics[width=0.5\textwidth]{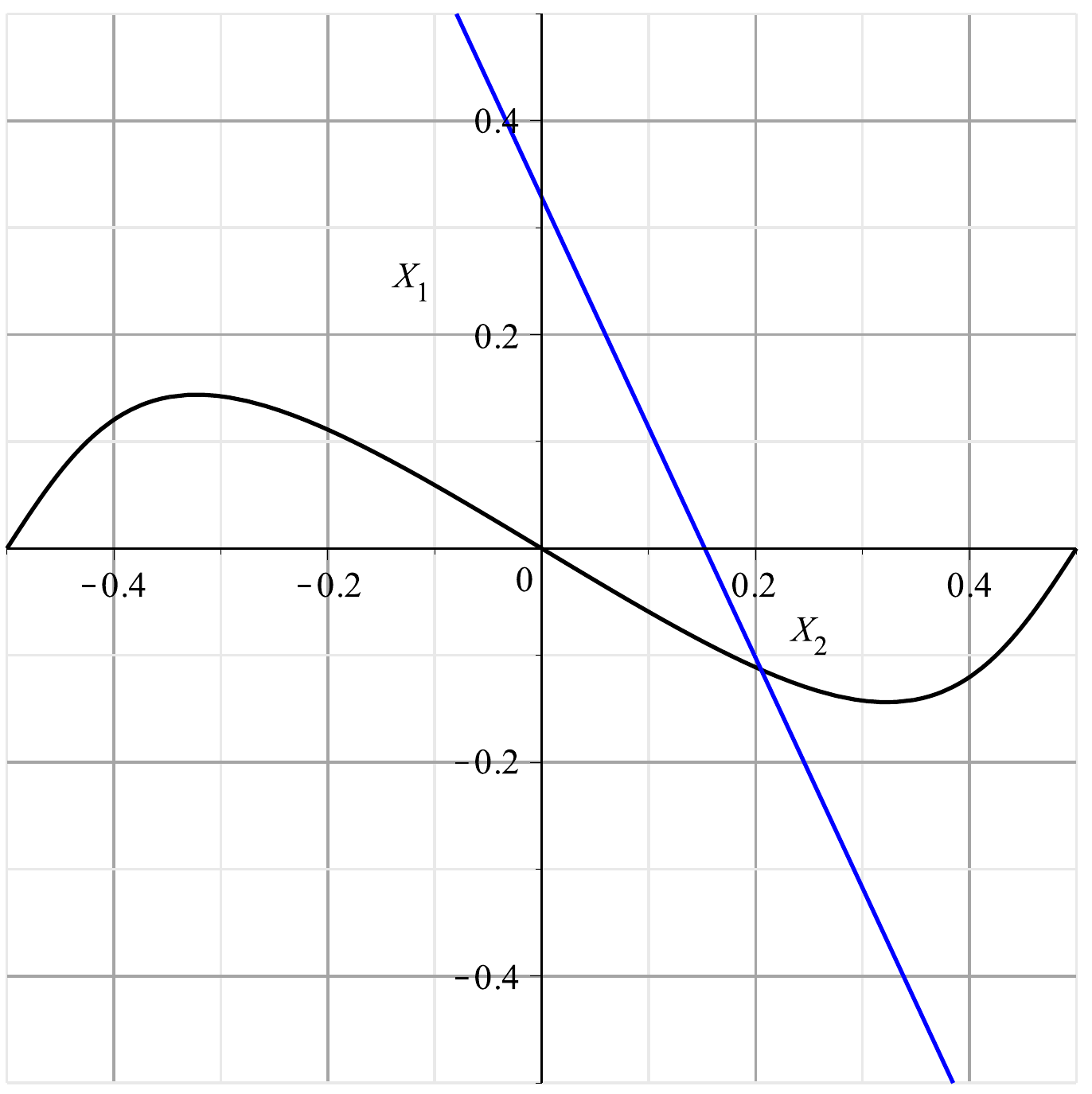}
\includegraphics[width=0.5\textwidth]{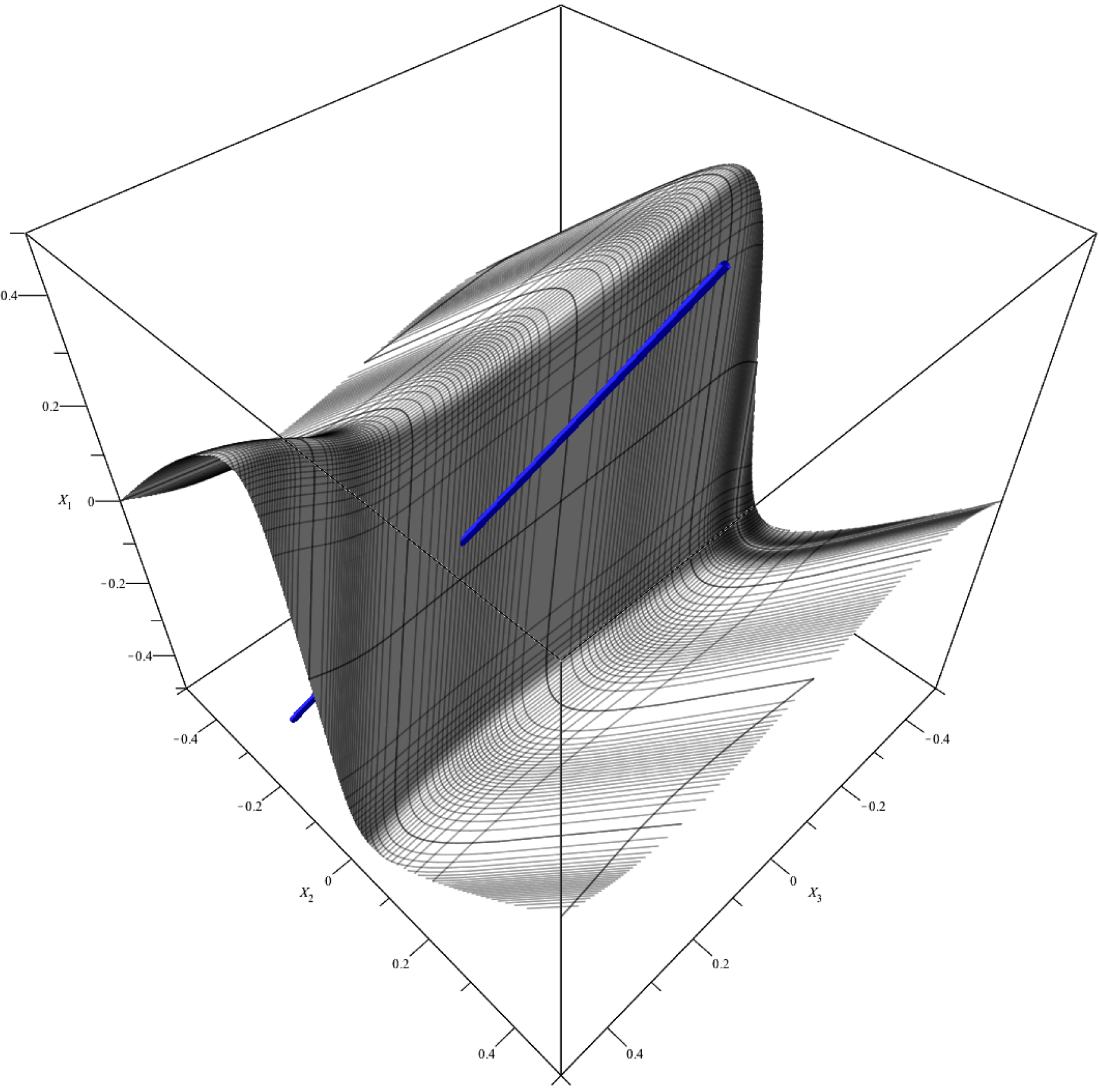}
\caption{Intersection of the line $ W(\k)-W_0$ (blue or lighter line) with the theta divisor  $(\Th)_R$ in $\mathbb T_g$,
where $g=2$ 
(left panel) and  $g=3$ 
(right panel). On the left panel ($g=2$) $(\Th)_R$ is represented by a curve, on the right panel ($g=3$) $(\Th)_R$ is represented by a surface.
In both cases the point of intersection  of $ W(\k)-W_0$ with $(\Th)_R$ determines some  $\k=\tk_n$.
}
 \label{ThetaDivisor}
\end{figure}

Moreover, since $(\Th)_R$ is $Z^g$ periodic on $\Pi$, it is sufficient to study  $(\Th)_R$ in a $g$ (real)
dimensional torus $\mathbb T_g$. Numerically simulated surfaces  $(\Th)_R\cap\mathbb T_g$ for $g=2,3$,
and their intersections with the line  $ W(\k)-W_0$ are shown on Figure \ref{ThetaDivisor}.
In the case $g=2$ we proved that the line $ W(\k)-W_0$ has one and only one intersection with  $(\Th)_R$ in $\mathbb T_2$.
It is likely (but not proven yet) that this statement holds  for a general $g\in\N$. However, the following lemma is sufficient to obtain
 the asymptotics \eqref{in1} for $\l_n$ with any $g\in\{2,3,\dots\}$.

\bl\label{prop-ass}
[Prop.7.11]
Let $\k_0\in\R^+$ and  $g\in\{2,3,\dots\}$. For any $N\in\N$ the number $m(N)$ of intersections of 
the segment of the line $W(\k)-W_0$, where $\k \in \le [\k_0 , \k_0 + \frac{N(g-1) i\pi}{\tau_{11}}\ri)$,
with $(\Theta)_\R$ is bounded by
\be\label{bound_n}
(N-1)(g-1)\leq m(N) \leq (N+1)(g-1).
\ee
\el

Let us now denote
\begin{equation}\label{some_not}
\begin{split}
&\mathbf f_n:= W(\tk_n)-W_0,~~~\gg(z) = \frac 12 - 2 \int_{a_1}^z\!\!\!  \omega_1d z ~~~{\rm and} \\
d(z)=&\frac{R(z)}{2\pi i}\le(-\sum_{j=1}^{g+1}\int_{a_{2j-1}}^{a_{2j}}\frac{\ln w(\z)d\z}{(\z-z)R_+(\z)}+
\sum_{j=1}^{g}\int_{a_{2j}}^{a_{2j+1}}
\frac
{i\d_{\mu(j)}
d\z}{(\z-z)R_+(\z)}\ri),
\end{split}
\end{equation}
where: $\mu(g)=0$ and $\mu(j)=j$ for all $j\neq g$; 
the vector $\vec\d=[\d_1,\dots,\d_{g-1},\d_0]^t$  is given by $\vec \d = 2\pi L^{-1}  \le(2\mathfrak u(\infty) - \mathfrak u(a_{2g+2}) \ri)$
and
\be
L=
\le[
\begin{array}{ccccc}
1 & 0 & \dots&0 & -1\\
0 & 1& 0\dots &0& -1\\
& &\ddots &\\
&&\dots &1&-1\\
0&0&\dots &0&1
\end{array}
\ri].
\ee

\bp\label{propositiongg}[Prop.4.2] 
{\bf (1)} $\gg(z)$ satisfies  the jump conditions
\be\label{geqm}
\gg_+(z)+\gg_-(z)=-1~~~~{\rm on}~I_i,~~~~~\gg_+(z)+\gg_-(z)=1~~~~{\rm on}~I_e,
\ee
\be\label{geqc}
~{\rm and}~~~~~\gg_+(z)-\gg_-(z)=i\O_{\m(j)}~~~~{\rm on}~[a_{2j},a_{2j+1}],~~j=1,\cdots,g,
\ee 
where 
$\O_0= \frac 4 i \sum_{k=1}^g\int_{a_{2k-1}}^{a_{2k}}\!\!\!\!\! \omega_1dz\in \R$ and
 $\Omega_j = \frac 4 i \sum_{k=1}^j\int_{a_{2k-1}}^{a_{2k}} \!\!\!\!\! \omega_1dz\in \R$.
{\bf (2)}
The function $d(z)$given by \eqref{some_not}
 is analytic on $\bar \C \setminus [a_1,a_{2g+2}]$ (in particular, analytic  at infinity) and satisfies the jump conditions 
\be
d_+  +d_- =- \ln w~~{\rm on~} I,~~~\qquad
 d_+  -d_- =i\d_{\m(j)}~~{\rm on~} c_j,~ ~[a_{2j},a_{2j+1}],~~j=1,\cdots,g.
\label{propertyDelta}
\ee
\ep

Let \be
r(z):=  \sqrt[4]{\frac {\prod_{j\in J} (z-a_j)}{\prod_{\ell\in J'} (z-a_\ell)}}, \ \ \ z\in \C \setminus [a_1,a_{2g+2}],
\label{spinorh}
\ee
where  $J = \{ 1,5, 7,9,11, \dots, 2g-1\}$ and $J' = \{1,2, 3, \dots 2g+2\}\setminus J$ (so that $|J| = g-1$ and $|J|' =g+3$). 
The function $r(z)$ is defined so that it is analytic in $\C\setminus [a_1,a_{2g+2}]$ and at infinity behaves like $\frac 1 z$.

Let 
\begin{equation}
\label{Ups}
\begin{split}
\Upsilon^{(j)}(z;\mathbf f_n ) = (-1)^j&\sqrt{ \frac {\Theta(W_0\!+\!(-1)^j2\mathfrak u(\infty) )}{\Theta(\mathbf f_n\!+\! (-1)^j2\mathfrak u(\infty))}
 \frac {[\mathbb A^{-1} \nabla \Theta(W_0)]_g }{i\vec \tau_1\!\!\cdot \!\! \nabla \Theta(\mathbf f_n) }}
 \\
&\times 
\frac{\Theta\le(\mathfrak u_+(z)\! +\!\!(-1)^j\!\mathfrak u(\infty) + \mathbf f_n \ri) r_+(z)}
{ \Theta\le(\mathfrak u_+(z) \!+\!(-1)^j \mathfrak u(\infty)+W_0\ri)},~j=1,2,
\end{split}
\end{equation}
where $z\in I$. It follows from Corollary 7.20, \cite{BKT2}, that for every $n\in\N$ we have  $\Upsilon^{(1)}(z;\mathbf f_n)\equiv \pm\Upsilon^{(2)}(z;\mathbf f_n)$, where the choice of the sign depends on a particular $n$. It turns out that this sign is not 
essential, since the normalized singular functions 
$\wh f_n(z)$ and $\wh h_n(z)$, approximated through $\Upsilon^{(j)}(z;\mathbf f_n )$ (see below),
  are determined only up to a sign. Thus, we introduce
$\Upsilon(z;\mathbf f_n)$ that, for a given $n\in\N$,  coincides with both 
$\Upsilon^{(j)}(z;\mathbf f_n)$, $~j=1,2,$ modulo factor $(-1)$.

Now the asymptotics of singular functions is described by the following theorem.

\bth \label{cor-first}[Thm.7.22]
The  singular functions $\wh f_n(z)$ and $\wh h_n(z)$ of the 
system in \eqref{svd-def2} {\em normalized} in $L^2(I_i)$ and $L^2(I_e)$, respectively,  are asymptotically given by
\begin{equation}
\label{843}
\begin{split}
\wh f_n(z) =
{i} \Im  \le[ 2\Upsilon(z;\mathbf f_n){\rm e}^{-i \tk_n \Im (\gg_+(z))  -i\Im (d_+(z)) } \ri] + 
\mathcal O({\tk}_n^{-1}), ~~~z\in I_i,\\
\wh  h_n(z) = 
\Re  \le[ 2\Upsilon(z;\mathbf f_n) {\rm e}^{-i \tk_n \Im (\gg_+(z))  -i\Im (d_+(z)) } \ri] + 
\mathcal O({\tk}_n^{-1}),~~~z\in I_e,
\end{split}
\end{equation}
where the approximation is uniform in any compact subset of the interior of $I_i,  I_e$, respectively.
%
\et

\bc \label{cor-sing-func}[Cor.7.24]
 The {singular functions} $f_n(z)$ and $h_n(z)$  of the system \eqref{svd-def} {\em normalized} in $L^2(I_i,\frac 1{w(z)})$ and $L^2(I_e,\frac 1{w(z)})$, respectively,  are asymptotically given by
\bea
\wh f_n(z) =
\sqrt{w(z)} \Im  \le[ 2\Upsilon(z;\mathbf f_n){\rm e}^{-i \tk_n \Im (\gg_+(z))  -i\Im (d_+(z)) } \ri] + 
\mathcal O({\tk}_n^{-1}), ~~~z\in I_i,\cr
\wh  h_n(z) = 
\sqrt{w(z)}\Re  \le[ 2\Upsilon(z;\mathbf f_n) {\rm e}^{-i \tk_n \Im (\gg_+(z))  -i\Im (d_+(z)) } \ri] + 
\mathcal O({\tk}_n^{-1}), ~~~z\in I_e,\label{844}
\eea
where the approximation is uniform in any compact subset of the interior of $I_i,  I_e$, respectively.
\ec

\section{Instability of the interior problem in Sobolev spaces}\label{sec-sobol}


\subsection{Continuation of $f$ from $I_i$}\label{sec-cont}

The function $\psi(y)$ in \eqref{int-eq} is analytic in $\C\setminus I_e$ and is known on $I_i$. If we can find the analytic continuation of $\psi(y)$ on $(a_2, a_{2g+1})$, then, according to \eqref{psi}, we can solve the problem of reconstructing $f$ on $(a_2, a_{2g+1})$. 

The idea of such reconstruction is straightforward. The eigenfunctions $\phi_n=\frac 1 {\sqrt{2}}(\wh f_n\chi_i + \wh h_n\chi_e)$ of the self-adjoint
Hilbert-Schmidt integral operator $\wh K: L^2(I)\mapsto  L^2(I)$ form an orthonormal basis in  $L^2(I)$. Thus, $\wh f_n, \wh h_n$ form 
orthonormal bases
in $L^2(I_i)$, $L^2(I_e)$ respectively, so that  $ f_n,  h_n$ form 
orthonormal bases in the corresponding
in $L^2(I_i,1/w)$, $L^2(I_e,1/w)$. Note that the former 
{coincides} with $L^2(I_i)$. Given $\psi\in L^2(I_i,1/w)$ and $\varphi\in L^2(I_e,1/w)$ we have
\be\label{psi_phi_exp}
\psi=\sum \psi_n  f_n ~~{\rm on}~~ I_i~~{\rm and}~~~ \varphi=\sum \varphi_n  h_n~~{\rm on}~~ I_e,
\ee
where $\sum \psi_n^2<\infty,~\sum \varphi_n^2<\infty$.
According to \eqref{svd-def2}, $\CH_e^{-1} h_n =2\l_n f_n$, so that $\CH_e^{-1} \varphi =\psi$ and \eqref{psi_phi_exp} imply
$\psi_n=2\l_n\varphi_n$. In view of the asymptotics \eqref{in1} of $\l_n$, we conclude that the coefficients $\psi_n$ decay
exponentially fast, so we have a very fast convergence of the series \eqref{psi_phi_exp} for $\psi$. 
Note that, according to \eqref{svd-def}, the singular functions $f_n$ are analytic in $\C\setminus I_e$.
Thus the question of analytic continuation of $\psi$ to  $(a_2, a_{2g+1})$ through the series \eqref{psi_phi_exp}
is reduced to the question of convergence of $\psi=\sum \psi_n  f_n$ in $(a_2, a_{2g+1})\setminus I_i$.

Let $\Iscr_\o$, $\o>0$, denote the set of all  $z\in (a_2, a_{2g+1}) \setminus I_i$ that are at least $\o$ away from the nearest
branchpoint $a_j$, $j=2,3,\dots,2g+1$. Below, we consider only such $\o$, that $a_j+\o<a_{j+1}-\o$  for all $j=2,\dots,2g$.

\bl\label{lem-f_n_in_gaps}
There exists a constant $C_\o>0$, such that for all
  $n\in\N$ and for all $z\in \Iscr_\o$
\be\label{est-f_n-gaps}
|f_n(z)|\leq C_\o e^{\k_n(\Re\gg(z)+\hf)}.
\ee
\el

Lemma \ref{lem-f_n_in_gaps} follows from Lemma \ref{lem-est-gaps}, \eqref{acc_sing_val} and  \eqref{svd-def2}.

\bl\label{lem-Re_g}
$|\Re \gg(z)|< \frac 12$  for any $z\in\C\setminus I$, with  $\Re \gg(z)\equiv \hf$ on $I_e$ and 
$\Re \gg(z)\equiv-\hf$ on $I_i$.
\el

\begin{proof}
Consider $\gg(z)$ on the main sheet of the Riemann surface  $\mathcal R$ with branchcuts on $I$. 
Note that $\gg(z)$ is Schwarz symmetrical and satisfies the jump conditions $\gg_+ + \gg_- \equiv 1 $
on $I_e$ and   $\gg_+ + \gg_- \equiv -1 $
on $I_i$, see Proposition \ref{propositiongg}. Thus,  $\Re \gg(z)\equiv \hf$ on $I_e$ and 
$\Re \gg(z)\equiv-\hf$ on $I_i$. The remaining statement follows from the maximal principle
for harmonic functions.
\end{proof}

\bth\label{thm33}
For a given $\o>0$, the series $\psi(z)=\sum \psi_n  f_n(z)$ converges absolutely and uniformly on $\Iscr_\o$. 
\et

\begin{proof} Recall that $\la_n=\exp(-\k_n)$. As a consequence of Lemma \ref{lem-f_n_in_gaps}, we have
\be\label{unif-conv}
\sum \le | \psi_n  f_n(z)\ri|  \leq 2C_\o\varphi_* \sum e^{\k_n(\Re\gg(z)-\hf)},
 \ee
where $\varphi_*=\max_n\{|\varphi_n|\}<\infty$. In light of \eqref{in1} and Lemma \ref{lem-Re_g},
the series in the right hand side of \eqref{unif-conv} converges absolutely and uniformly on 
$\Iscr_\o$.
\end{proof}

\bc\label{analcont-series}
The series $\psi(z)=\sum \psi_n  f_n(z)$ provides analytic continuation of $\psi$ onto  $(a_2,a_{2g+1})$.
\ec
Indeed, by choosing a sufficiently small $\o$, one can analytically continue $\psi(z)$ to any point in  
$(a_2,a_{2g+1})\setminus I_i$
through this series.

\subsection{Instability of analytic continuation in Sobolev norms}

In the previous section we obtained a formula for analytic continuation of $\psi(y)$ from $I_i$ to all of $(a_2,a_{2g+1})$. Next we prove that analytic continuation of $\psi$ from $I_i$ is unstable for any pair of Sobolev spaces: $H^{s_1}(I_i)\to H^{-s_2}({\bf J})$, where ${\bf J}$ is any open set containing $I_i$. Clearly, it makes sense to consider $s_1,s_2>0$. For simplicity we will assume that $s_1$ and $s_2$ are integers, so (see Chapter 1 in \cite{egs}):
\begin{equation}\label{norm1}
\lVert f \rVert_{H^{s_1}(I_i)}^2:=\sum_{j=0}^{s_1} \int_{I_i} |f^{(j)}(y)|^2dy,
\end{equation}
and
\begin{equation}\label{norm2}
\lVert f\rVert_{H^{-s_2}(J)}:=\inf_{\tilde f \in H^{-s_2}(\mathbb R), f=\tilde f|_{\bf J}}\sup_{\phi\in\coi({\mathbb R})} \frac{\left| \int_{\mathbb R} \tilde f(y)\overline{\phi(y)}dy\right|}{\lVert \phi\rVert_{H^{s_2}({\mathbb R})}}.
\end{equation}
Let $\gamma$ be a collection of simple loops in the complex plane so that $I_i$ is  contained in the union of the interiors
of the loops. We take $\g$ to be  sufficiently close to $I_i$. By the Cauchy integral theorem using the analyticity of $f_n$ one can show that 
\begin{equation}\label{est1}
\lVert f_n\rVert_{H^{s_1}(I_i)} \leq c(s_1,\ga) \max_{z\in\gamma}|f_n(z)|
\end{equation}
for some $c(s_1,\ga)>0$. Analogously to Lemma \ref{lem-f_n_in_gaps}, it follows from Lemma \ref{lem-est-gaps} that 
\begin{equation}\label{est2}
\max_{z\in\gamma}|f_n(z)|\leq c_\ga \exp(\k_n(\max_{z\in\ga} \Re g(z)+\frac12))
\end{equation}
for some $c_\ga>0$. By taking $\ga$ sufficiently close to $I_i$, we can make $\max_{z\in\ga} \Re g(z)+\frac12$ as close to zero as we want.

\bl\label{lem-seq-intervals}
One can find a sequence of intervals $J_n\subset {\bf J}$ with the following properties:
\begin{enumerate}
\item The length of each $J_n$ is greater than a fixed positive constant independent of $n$;
\item The distance of each $J_n$ to $I_i$ is greater than a fixed positive constant independent of $n$; and
\item There exists $N>0$ large enough such that
\begin{equation}\label{est4}
|f_n(y)|\ge c \exp(\k_n(\Re g(y)+\frac12)),\ n\ge N,\ y\in J_n,
\end{equation}
for some $c>0$ independent of $n$. 
\end{enumerate}
\el

Lemma~\ref{lem-seq-intervals}  is proven in Appendix~\ref{proofstechn}. By property 1 in Lemma~\ref{lem-seq-intervals} we can find $L>0$ such that the length of each interval $J_n$ is greater than or equal to $L$. Then we select a real-valued function $\phi\in C_0^\infty([-L/2,L/2])$, $\phi\ge0$, $\phi\not\equiv0$. By shifting $\phi$ appropriately, we get a collection of functions $\phi_n\in\coi(J_n)$ and they all have the same  $H^{s_2}({\mathbb R})$-norm. Using the facts that: (i) $f$ and $\tilde f$ coincide on ${\bf J}$ (cf. \eqref{norm2}); (ii) $f_n$'s are real-valued on ${\bf J}$, and; (iii) $f_n$'s do not change sign on $J_n$ for $n$ large (cf. \eqref{est4}), equation \eqref{norm2} immediately yields
\begin{equation}\label{est3}
\lVert f_n \rVert_{H^{-s_2}({\bf J})}\ge c_\phi \min_{y\in J_n} |f_n(y)|,\ n\ge N,
\end{equation}
for some $c_\phi>0$. 

From the second property in Lemma~\ref{lem-seq-intervals}, by choosing $\ga$ sufficiently close to $I_i$ so that all $J_n$ are in the exterior of $\gamma$ and $\text{dist}(\ga,\cup_n J_n)>0$, we get $\inf_{y\in \cup J_n} \Re g(y) > \max_{z\in\ga} \Re g(z)$. Hence,
\begin{equation}\label{final}
\frac{ \exp(\k_n(\min_{y\in J_n} \Re g(y)+\frac12))}{ \exp(\k_n(\max_{z\in\ga} \Re g(z)+\frac12))}\to\infty,\ n\to\infty.
\end{equation}
Hence it follows from \eqref{est2} and \eqref{est4} that the quantity $\lVert f_n \rVert_{H^{-s_2}({\bf J})}$ cannot be bounded in terms of $\lVert f_n \rVert_{H^{s_1}(I_i)}$. Since the Sobolev norm $\lVert f \rVert_{H^s}$ is a monotonically increasing function of $s$ (provided that $f$ belongs to the appropriate spaces), our argument proves the following result.

\bth\label{instability} Fix an open set ${\bf J}\supset I_i$. The operation of analytic continuation from $I_i$ to ${\bf J}$ described in Corollary \ref{analcont-series} cannot be extended to a continuous operator $H^{s_1}(I_i)\to H^{-s_2}({\bf J})$ for any $s_1,s_2$.
\et

Theorem~\ref{instability} shows that analytic continuation is more unstable than calculation of any number of derivatives. An interesting question is to estimate the degree of ill-posedness of analytic continuation. This can be done, for example, by finding a Hilbert space $\mathcal A$  on which the operator of analytic continuation is bounded. It is clear that the space $\mathcal A$ should contain at least all functions in the range of $\CH^{-1}_e:\ L^2(I_e,1/w)\to L^2(I_i,1/w)$. If $\psi\in \mathcal A$, but $\psi$ is not in the range of $\CH^{-1}_e$, then the analytic continuation of $\psi$ is understood via the summation of the series in Corollary~\ref{analcont-series}.

Let $w_n$ be a sequence of positive numbers. Introduce the following space:
\begin{equation}
\mathcal A:=\{ \psi\in {L^2}(I_i):\, \sum_{n\ge0} w_n^2 |\psi_n|^2<\infty \},
\end{equation}
where 
\begin{equation}
\psi_n:=\langle \psi,f_n \rangle:=\int_{I_i} \psi(y) f_n(y) \frac1{w(y)}dy.
\end{equation}
It is obvious that $\mathcal A$ is a Hilbert space with the inner product defined by the formula
\begin{equation}
\langle \psi^{(1)},\psi^{(2)} \rangle:=\sum_{n\ge 0} w_n^2 \psi^{(1)}_n \overline{\psi^{(2)}_n}.
\end{equation}
\bth\label{stability} Fix an open set ${\bf J}$, whose closure is a subset of $(a_2,a_{2g+1})$. Suppose that each connected component of ${\bf J}$ contains at least one of the intervals that make up $I_i$. Suppose the sequence of $w_n$'s is such that the limit below exists and satisfies
\begin{equation}
0<\lim_{n \to \infty} \left\{\frac{w_n}n \exp(-\k_n(\sup_{z\in {\bf J}} \Re g(z)+\frac12))\right\} <\infty.
\end{equation}
Then one has: (1)  $\CH^{-1}_e(L^2(I_e,1/w))\subset \mathcal A$, and; (2) the operator of analytic continuation acting between the spaces $\mathcal A \to L^2({\bf J})$ is continuous.
\et
\begin{proof} 
Similarly to the proof of Theorem~\ref{thm33}, it is easily seen that assertion (1) holds. Now we prove assertion (2). First we  show that 
\begin{equation}\label{est-comb}
\max_{z\in {\bf J}}|f_n(z)|\leq c_ {\bf J} \exp(\k_n(\sup_{z\in {\bf J}} \Re g(z)+\frac12))
\end{equation}
for some $c_ {\bf J}>0$. Denote $G:=\sup_{z\in {\bf J}} \Re g(z)$. Let $\ga$ be a collection of simple contours in $\C$ containing the components of ${\bf J}\cap I_i$ in their interior. By making $\ga$ as close to these component as we need and using Lemma~\ref{lem-Re_g}, we can find $\ga$ such that $\sup_{z\in \ga} \Re g(z)<G$. Now \eqref{est-comb} follows immediately by using inequalities \eqref{est-f_n-gaps} and \eqref{est2} combined with the maximum modulus principle. Finally, to prove (2) we fix any $N>0$. Then
\begin{equation}\label{eq316}
\begin{split}
\int_{J} \left| \sum_{n=0}^N \psi_n f_n(z) \right|^2 dz \leq
|J| \left (\sum_{n=0}^N |\psi_n| \sup_{z\in J} |f_n(z)| \right)^2\leq
c \left (\sum_{n=0}^N |\psi_n| \frac{w_n}{n} \right)^2\leq
c \sum_{n=0}^N (|\psi_n| w_n)^2 \sum_{n=0}^N \frac{1}{n^2},
\end{split}
\end{equation}
where $c>0$ is some constant. By taking the limit $N\to\infty$ the desired assertion follows immediately.
\end{proof}
\begin{remark} Using the fact that the singular functions $f_n$ are analytic on ${\bf J}$ and the coefficients $\psi_n$ go to zero sufficiently fast, similarly to the proof of Theorem~\ref{thm33} and \eqref{eq316} it is easy to see that each $\psi\in\mathcal A$ defined on ${\bf J}$ via the series in Corollary~\ref{analcont-series} is a uniform limit of analytic functions. Hence the continuation of $\psi$ from $I_i$ to ${\bf J}$ via the series and via the conventional analytic continuation coincide.
\end{remark}


\section{Approximation in $L^2(I_i)$ }\label{sec-L2-conv}

According to Theorem \ref{cor-first}, the normalized singular functions $\wh f_n$ are approximated by 
\be
\tf_n:={i} \Im  \le[ 2\Upsilon(z;\mathbf f_n){\rm e}^{-i \k_n \Im (\gg_+(z))  -i\Im (d_+(z)) } \ri]
\label{tfn}
\ee
with accuracy $O(n^{-1})$ in the sup-norm (uniformly) on any compact subset of the interior of $I_i$. 
In this subsection 
 we discuss this approximation in $L^2(I_i)$.
We will use 
$\|f\|$ to denote the
$L^2$ norm of $f\in L^2(I_i)$.

\bl\label{lem-prepar}
Let $\o_0$ be so small that each interval $(a_k-\o_0,a_k+\o_0)$ contains no endpoints except $a_k$.
Then there exists some $\eta>0$, such that 
\be\label{est-unif}
\forall k\in\{3,\dots,2g\},\,\forall n\in\N,\, \forall \o\in(0,\o_0):~~~ |\tf_n(z)|\leq \frac{\eta}{|z-a_k|^\frac 14} ~~{\rm on}~~(a_k-\o,a_k+\o).
\ee
\el

Lemma \ref{lem-prepar} follows from Lemma \ref{lemmab1} and  Corollary \ref{cor-Ups12}.

\bl\label{lem-norm-tf}
{The norms of $\wt f_n(z)$ \eqref{tfn} satisfy the asymptotic expansion}
\be\label{norm-tf-st}
\, \|\tf_n\|=1 + O(n^{-1})\ ,\ \ n\to \infty.
\ee
\el


The proof of this lemma can be found in Appendix \ref{proofstechn}.

Let $\o>0$ and define 
$I_i^\o=I_i\setminus \bigcup_{k=3}^{2g}(a_k-\o,a_k+\o)$. If $f\in L^2(I_i)$, then $\|f\|^2= \|f\|_b^2+\|f\|^2_t$,
where $\|f\|_b$ denotes the norm of $f$ in  $L^2(I_i^\o)$ (in the bulk) and  $\|f\|_t$ denotes the norm of $f$ in  $L^2(I_i\setminus I_i^\o)$
(in the tails).

According to Theorem \ref{cor-first}, for any $\o\in(0,\o_0)$ there exists some $P_\o>0$, such that 
\be\label{dif-norm}
\|\wh f_n-\tf_n\|_b\leq \frac{P_\o}{n}.
\ee

\bth
$\tf_n$ approximate $\wh f_n$ in $L^2(I_i)$, that is, $\forall \e>0 \, \exists n_0\in\N$ such that 
$ \forall n>n_0:\,\|\wh f_n-\tf_n\|<\e$. 
\et 
\begin{proof}
 According to \eqref{est-unif}, $\|\tf_n\|_t\leq 2\sqrt{g-1}\eta\o^\frac 14$ for all $n\in\N$.
As implied by Lemma \ref{lem-norm-tf}, there exist some $Q_\o>0$, such that $\|\tf_n\|\geq 1- \frac{Q_\o}{n}$.
Since $1- \frac{Q_\o}{n}\leq\|\tf_n\|\leq \|\tf_n\|_b+\|\tf_n\|_t$, we obtain
$\|\tf_n\|_b\geq 1-2\sqrt{g-1}\eta\o^\frac 14- \frac{Q_\o}{n}$. Then, according to \eqref{dif-norm},
\be
\begin{split}
& \|\wh f_n\|_b\geq \|\tf_n\|_b-\|\wh f_n-\tf_n\|_b\geq  1-2\sqrt{g-1}\eta\o^\frac 14-\frac{P_\o}{n}- \frac{Q_\o}{n}, ~~~{\rm so ~that} \cr
& \|\wh f_n\|_t^2=1- \|\wh f_n\|_b^2 \leq 2\le(2\sqrt{g-1}\eta\o^\frac 14+\frac{P_\o+Q_\o}{n}\ri).
\end{split}
\ee
Thus,
\be\label{first-est}
\|\wh f_n-\tf_n\|\leq \|\wh f_n-\tf_n\|_b +\|\wh f_n\|_t+ \|\tf_n\|_t\leq 2\sqrt{g-1}\eta\o^\frac 14+\frac{P_\o+Q_\o}{n}+\sqrt{2\le(2\sqrt{g-1}\eta\o^\frac 14+
\frac{P_\o+Q_\o}{n}\ri)}.
\ee
It is clear that for a small $\e$ condition $2\sqrt{g-1}\eta\o^\frac 14+\frac{P_\o+Q_\o}{n}<\frac{\e^2}{4}$ would imply $\|\wh f_n-\tf_n\|\leq \e$.
 Choose $\o^\frac 14=\frac{\e^2}{16\sqrt{g-1}\eta}$. 
Then the former inequality holds for all $n>\frac{8(P_\o+Q_\o)}{\e^2}$. The proof is completed.
\end{proof}

\appendix
\section{Approximate solution of the RHP \ref{RHPGamma} and related results from \cite{BKT2}}\label{sec-ideas}

Construction of the leading order approximation of the solution $\G(z;\l)$ of the RHP \ref{RHPGamma} in the limit $\l\ra 0^+$ 
is at the heart of our method. We also have to control the accuracy of such approximation. We employ the nonlinear steepest
descent method of Deift and Zhou, that allows to asymptotically reduce the original RHP (RHP \ref{RHPGamma}) to a certain RHP 
with constant jumps (RHP \ref{modelRHP})   
that one can solve explicitly. The asymptotic reduction consists of a sequence of transformations
of the RHP \ref{RHPGamma}, some of them equivalent and some asymptotic (with the error estimates for the later). 
The key idea is a factorization of the jump matrix
with a subsequent contour deformation, where each factor ``aquiring'' its own jump-contour in the process. In this appendix we 
only briefly outline some main points of the reduction of the  RHP \ref{RHPGamma} and provide a solution to the 
corresponding ``reduced'' RHP with constant jumps. The details can be found in \cite{BKT2}. 
There exists a large and rapidly growing literature about
 the  method Deift and Zhou and its various applications, see, for example, \cite{Deift}, \cite{Deift60volume}.
We also include some facts about theta divisors as well as some further results  from  \cite{BKT2}
that are used in the proof of technical lemmas in Appendix \ref{proofstechn}.

Let $\Si$ be an 
oriented collection of contours that partition $\C$ into a finite number of open
regions and let $V(z)$ be an $n\times n$ matrix valued function  defined on $\Si$, satisfying
certain conditions at the nodes of  $\Si$. \footnote{A point $z\in\Si$ is called a node if three or more branches of $\Si$
emanates from $z$. We assume  $\Si$  has no more than finitely many nodes. }
A (somewhat) general formulation of a matrix RHP can be stated as follows.  We do not get here into the details of the smoothness
of $\Si$ and $V(z)$.

\begin{problem}
\label{RHPgen}
Find an $n\times n$ matrix-function $M(z)$ that:
\bi
\item is analytic in each element of partition, induced by the contour  $\Si$;
\item  for any $z\in\Si$ that is not a node $M(z)$ admits non-tangential boundary values $M_\pm(z)$ from the corresponding
sides of $\Si$ and
\be\label{jump-gen}
M_+(z)=M_-(z)V(z);
\ee
\item
\be
\label{ass-gen}
\lim_{z\ra\infty} M(z)=\1.
\ee
\ei
\end{problem}

In general, the existence of a solution to the RHP \eqref{RHPgen} is not guaranteed.
The nonlinear steepest descent method  is based upon the following ``small norm theorem''.

\bth\label{theo-small_norm}
Let $N_p$ denotes the norms of $V(z)-\1$ in $L^p(\Si,dz)$. Then
\bi
\item There is a constant $C_\Si$  such that if $N_\infty < C^{-1}_\Si$ the solution of the RHP \ref{RHPgen} exists;
\item In this case 
\be\label{small_norm}
\| M(z)-\1\|\leq \frac{1}{2\pi {\rm dist}(z, \Si)}\le(N_1+\frac{C_\Si N_2^2}{1-C_\Si N_\infty}\ri)
\ee
for every $z\in\C\setminus \Si$.
\ei
\et

The name of this theorem reflects the fact 
 that 
the solution $M(z)$ of the RHP \ref{RHPgen} 
is  close (pointwise)  to the
identity matrix $\1$ {if the norms $N_{1,2}$ are small}.

Let $\k=-\ln \la$.  Then $\k>0$ when $\la\in(0,1)$ and $\k\ra\infty$ as $\l\ra 0$. The first transformation
is replacing $\G(z;\l)$ with $Y(z;\k)$ by 
\begin{equation}\label{y-def}
Y(z;\k)=e^{-(\varkappa \gg(\infty)+d(\infty)\sigma_3}\Gamma(z;e^{-\k}) e^{(\varkappa \gg(z)+d(z))\sigma_3},
\end{equation}
where $ \gg(z),d(z)$ are defined by \eqref{some_not} and the Pauli matrices are defined as
$$
\s_1= \begin{bmatrix}
0 & 1\\
1&0
\end{bmatrix},~~~
\s_2= \begin{bmatrix}
0 & -i\\
i&0
\end{bmatrix},~~~
\s_3= \begin{bmatrix}
1 & 0\\
0&-1
\end{bmatrix}.
$$

Then direct calculations show that the RHP \ref{RHPGamma} for $\G(z;\l)$ is reduced to the following equivalent RHP for $Y$.

\begin{problem}\label{prob-Y}
Find a $2\times 2$ matrix-function $Y(z;\k)$ with the following properties:
\begin{enumerate}
\item[{\bf (a)}] $Y(z;\k)$ is analytic in $\C\setminus[a_1,a_{2g+2}]$; 
\item[{\bf (b)}] $Y(z;\k)$ satisfies the jump conditions
\bea
\label{jumpY}
Y_+&\& =Y_- \left[
\begin{matrix}
 e^{(\k \gg +d)_+ -(\k\gg +d)_- } & 0 \\ 
{iw} e^{\k(\mathcal \gg_+ +\gg_- +1) + d_+ +d_-} & e^{-(\k \gg +d)_+ +(\k\gg +d)_- } 
\end{matrix}\right],  \ \ z\in  I_i,
\nonumber \\
Y_+&\&=Y_-\left[\begin{matrix} e^{(\k \gg +d)_+ -(\k\gg +d)_- } & 
 -\frac iw e^{-\k(\mathcal \gg_+ + \gg_- -1) - d_+-d_-  } \\ 0 & e^{-(\k \gg +d)_+ +(\k\gg +d)_- }\end{matrix}\right],\ \ z\in  I_e 
 \nonumber \\
\rm{and}~~ &\& Y_+=Y_- e^{[(\k\gg+d)_+ -(\k\gg+d)_-] \sigma_3}~~{\rm on}~~ [a_{2j},a_{2j+1}],~ j =1,\dots, g ;
\eea
\item[{\bf (c)}]
$Y=\1+O(z^{-1})~~~~{\rm as} ~~z\ra\infty,$
 and;
\item[{\bf (d)}] Near the branchpoints (we indicate the behavior for the columns if these have different behaviors)
\be\label{endpcondY}
Y(z;\k)=[ \mathcal O(1), \mathcal O(z-a_{j})^{-\frac 1 2}],\ j=1,2g+2;
~~~~~ Y(z;\k)=O(\ln(z-a_j)), ~~~j=2,\dots, 2g+1.
\ee
\end{enumerate}
\end{problem}


In the next transformation (now of the RHP \ref{prob-Y}) we first factorize the triangular jump matrices
in \eqref{jumpY} as
\be
\label{jumpY2}
\begin{split}
Y_+&=Y_- 
\le[\begin{matrix}
1 & \frac { {\rm e}^{-\k (2\gg_{_-} +1)-2d_- }}{iw}\\
0 & 1
\end{matrix}\ri]
\le[\begin{matrix}
0& i\\ 
 i & 0
\end{matrix}\ri]
\le[\begin{matrix}
1 & \frac { {\rm e}^{-\k (2\gg_{_+} +1)-2d_+}}{iw}\\
0 & 1
\end{matrix}\ri]
\text{ on } I_i,\\
Y_+&=Y_-\le[\begin{matrix}
1 & 0\\
 {iw}{ {\rm e}^{ \k (2\gg_{_-} -1)+2d_-}} & 1
\end{matrix}\ri]
\le[\begin{matrix}
0& -i \\ 
-i& 0
\end{matrix}\ri]
\le[\begin{matrix}
1 & 0\\
{iw} { {\rm e}^{ \k (2\gg_{_+} - 1)+2d_+ }} & 1     
\end{matrix}\ri]
\text{ on } I_e,
\end{split}
\ee
and then put each of the three factors (for $I_e$ and for $I_i$) on its own jump contour as described below.

\begin{figure}
\begin{center}
\resizebox{0.7\textwidth}{!}{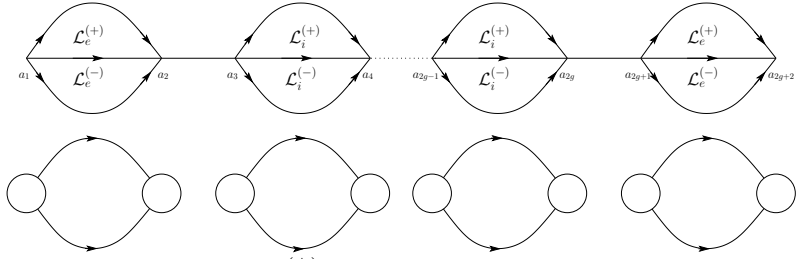}
\end{center}
\vspace{-20pt}
\caption{The regions of the lenses $\mathcal L^{(\pm)}_{i,e}$ (above) and the jumps of the error matrix $\mathcal E$ (below).}
\label{Lenses}
\end{figure}

The validity of the factorization can be checked directly, taking into the account the identities
$-\k(\gg_+ +\gg_- \pm 1)-\ln w - d_+ - d_- \equiv 0$ that hold on $I_i$ and $I_e$ respectively, see
\eqref{geqm}, \eqref{propertyDelta}. 
The left and right (triangular) matrices in both factorizations  (on $I_i$ and on $I_e$) admit analytic extension 
on the left/right vicinities of the corresponding
segments  because they are boundary values of analytic matrices in those vicinities. This suggests 
opening of  the lenses $\pa \mathcal L_{e}^{(\pm)},~ \pa \mathcal L_{i}^{(\pm)}$ around the corresponding intervals
of $I_e\cup I_i$,
see Figure \ref{Lenses} upper  panel,
and introduction of the new unknown matrix
\bea
Z(z;\k) =
\le\{\begin{array}{ll}
 Y(z;\k) & \text{ outside the lenses,}  \\
\ds  Y(z;\k) \le[\begin{matrix}
1 & 0\\
\mp {iw}{ {\rm e}^{ \k (2\gg -1)+2d}} & 1
\end{matrix}\ri] & z\in {\mathcal L_e^{(\pm)}},\\
\ds  Y(z;\k) \le[\begin{matrix}
1 & \frac{\mp 1}  {iw}{ {\rm e}^{ -\k (2\gg +1)-2d}} \\
0& 1
\end{matrix}\ri] & z\in  \mathcal L_i^{(\pm)}.
 \end{array}
 \ri.
 \label{422}
\eea

Consequently, after the second (equivalent) transformation we obtain  the following RHP for  the matrix $Z$.
\begin{problem}
\label{ZetaRHP}
Find the matrix $Z$, analytic on the complement of the arcs of Figure \ref{Lenses}, satisfying  the jump conditions (note also the orientations marked in Figure \ref{Lenses})
\bea
\label{RHPZ}
Z_+(z;\k) = Z_-(z;\k) \le\{
\begin{array}{ll}
{\rm e}^{i(\k \Omega_{\m(j)} + \d_{\m(j)})\s_3} & z\in[a_{2j},a_{2j+1}],~ j =1,\dots, g, \\
 \le[\begin{matrix}
1 & 0\\
{iw}{ {\rm e}^{ \k (2\gg -1)+2d}} & 1
\end{matrix}\ri]  & z\in  \pa \mathcal L_{e}^{(\pm)} \setminus \R,\\
\le[\begin{matrix}
1 & \frac{1}  {iw}{ {\rm e}^{ -\k (2\gg +1)-2d}} \\
0& 1
\end{matrix}\ri]  & z\in \pa \mathcal L_{i}^{(\pm)} \setminus \R,\\
i\s_1 & z\in I_i,\\
-i\s_1 & z\in I_e,
\end{array}
\ri.
\eea
normalized by 
\be
Z(z;\k) \to \1\ ,\ \ z\to \infty, \label{ass-Z}
\ee
and with the same endpoint behavior   as $Y$   near the endpoints $a_j$'s, see \eqref{endpcondY}. 
Here 
$$\vec\d=[\d_1,\dots,\d_{g-1},\d_0]^t=2\pi L^{-1}\le((2\mathfrak u(\infty) - \mathfrak u(a_{2g+2}) \ri),  ~~~
 \vec\O=[\O_1,\dots,\O_{g-1},\O_0]^t=-2iL^{-1}\t_1$$ 
and $\mu(g)=0$, $\mu(j)=j$ for all $j\neq g$, see Prposition \ref{propositiongg}.
\end{problem}


In the third and final transformation we would like to (asymptotically) reduce the RHP \ref{ZetaRHP} for $Z(z;\k)$
to the following RHP for $\Psi=\Psi(z;\k)$.

\begin{problem}[Model problem]
\label{modelRHP}
Find a matrix $\Psi=\Psi(z;\k)$, analytic on $\C\setminus[a_1,a_{2g+2}]$ and satisfying the following conditions:
\bea\label{jumpPsi}
& \Psi_+=\Psi_-  (-1)^{s(j)}(i\s_1), & z\in[a_{2j-1},a_{2j}],~~
j=1,\dots g+1,\cr
&\Psi_+=\Psi_-e^{i(\k\O_{\m(j)}+\d_{\m(j)})\sigma_3}  & z\in[a_{2j},a_{2j+1}],~ j =1,\dots, g, \cr
&\Psi(z) = \mathcal O( |z-a_j|^{-\frac 1 4}) \ , & z\to a_j, \ j=1,\dots, 2g+2,\cr
&\Psi(z) = \1 + \mathcal O(z^{-1})\ , & z\to\infty,~~{\rm and}~~\Psi_\pm(z)\in L^2([a_1,a_{2g+2}]).
\eea
Here $s(j) = \delta_{j,1}  + \delta_{j,g+1}$, where $\d_{j,k}$ denotes the Kronecker delta. 
\end{problem}

The RHP \ref{modelRHP} is not eqiuvalent to the RHP \ref{ZetaRHP} since the former does not have jumps
on the lenses  $\pa \mathcal L_{e}^{(\pm)} \setminus \R$, $\pa \mathcal L_{i}^{(\pm)} \setminus \R$.
Also,  a different behavior is required near the branchpoints $a_j$, $j=1,2,\dots,,2g+2$.
However, as a consequence of Theorem \ref{theo-small_norm}, the solution $\Psi(z;\k)$ of  the RHP \ref{modelRHP}
will approximate  the solution $Z(z;\k)$  of  the RHP \ref{ZetaRHP}, if the jump matrices on the lenses 
 $\pa \mathcal L_{e}^{(\pm)} \setminus \R$, $\pa \mathcal L_{i}^{(\pm)} \setminus \R$ in \eqref{RHPZ}
will be small in the norms $N_1,N_2$ and $N_\infty$.
The choice of
the so-called $g$-function $\gg(z)$  in the transformation \eqref{y-def}, as well as of the lenses 
$\mathcal L_{e,i}^{(\pm)}$, is defined
by the requirements that 
\be
\Re (2\gg -1) <0~~{\rm on}~~  \pa \mathcal L_{e}^{(\pm)} \setminus \R~~{\rm and}~~
\Re (2\gg +1) >0~~{\rm on}~~  \pa \mathcal L_{i}^{(\pm)} \setminus \R. 
\ee
If these requirements hold, 
 the jump matrices \eqref{RHPZ} on the contours  $\pa \mathcal L_{e}^{(\pm)} \setminus \R,
 \pa \mathcal L_{i}^{(\pm)} \setminus \R$ approach $\1$ exponentially fast  as $\k\ra\infty$  {in any $L^p,  \ p<\infty$ but {\em not} in $L^\infty$},
 { because}
this convergence is uniform 
away from small vicinities of the branchpoints. These vicinities  which require special consideration. 
Namely, to match RHP \ref{RHPZ} with RHP \ref{modelRHP},   we need to construct special {\em local parametrices} 
in these vicinities  of 
the branchpoints.

The discrepancy
between $Z(z;\k)$ and  $\Psi(z;\k)$, the latter modified by the parametrices near the branchpoints,   
is represented by the so-called error matrix $\mathcal E(z;\k)$. The error matrix also satisfies 
a certain RHP; the jump contours of this RHP are shown on Figure \ref{Lenses}. We already know
that the jump matrices on the arcs $\pa \mathcal L_{e}^{(\pm)} \setminus \R,
 \pa \mathcal L_{i}^{(\pm)} \setminus \R$ away from the branchpoints should approach $\1$ exponentially
fast in $\k\ra\infty$. The parametrices, constructed in \cite{BKT2}, ensure that
the jump matrices on the circles, shown on Figure \ref{Lenses}, behave like 
 $\1+O(\k^{-1})$ as $\k\ra\infty$ { in any $L^p$, including $L^\infty$}. Thus, according to Theorem \ref{theo-small_norm}, $\Psi(z;\k)$
is $O(\k^{-1})$ {close} to $Z(z;\k)$ uniformly in $z$ outside small circles around the branchpoints. 

Now, by reversing the chain of transformations, one obtains the following summary of the  steepest descent analysis of \cite{BKT2}:
let constants $\epsilon,  \rho_z, \rho_0>0$ be fixed and sufficiently small. Then 
\be
\label{gammaexact}
\Gamma (z;{\rm e}^{-\k}) = e^{(\varkappa \gg(\infty)+d(\infty))\sigma_3}
\mathcal E (z;\k)\Psi(z;\k) 
\le[\begin{matrix}
1 &  \frac{\pm 1}  {iw}{ {\rm e}^{ -\k (2\gg +1)-2d}}\wh\chi_{i,\pm}  \\
\pm {iw}{ {\rm e}^{ \k (2\gg -1)+2d}}\wh\chi_{e,\pm} & 1
\end{matrix}\ri] 
e^{-(\varkappa \gg(z)+d(z))\sigma_3}, 
\ee 
where $\wh\chi_{e}^\pm, \wh\chi_{i}^\pm$ are the characteristic (indicator) functions of the sets 
$\mathcal L_{e}^{(\pm)}$, $\mathcal L_{i}^{(\pm)}$ respectively, see Figure \ref{Lenses},  and  
\be\label{err_est}
\mathcal E(z;\k) = \1 + \frac {\mathcal O(\k^{-1})}{1+|z|}
\ee
 uniformly  in the domain 
\be\label{domain}
\Im \k<\epsilon,\ \ \ |\Theta\le(W(\k)-W_0 \ri)|>\rho_0,~~~
|z-a_j|>\rho_z\ ,\ \  j=1,\dots, 2g+2.  
\ee
%
The matrix $\mathcal E(z;\k)$ solves an auxiliary RHP where all the jumps satisfy the assumption of Theorem \ref{theo-small_norm}; in particular it is important for us that $\mathcal E$ does not have a jump on the main arcs $I_e\cup I_i$. 
 This implies that the following matrix 
\be
\label{gammainfty}
\Gamma^{(\infty)} (z;{\rm e}^{-\k}) = e^{(\varkappa \gg(\infty)+d(\infty))\sigma_3}
\Psi(z;\k) 
\le[\begin{matrix}
1 &  \frac{\pm 1}  {iw}{ {\rm e}^{ -\k (2\gg +1)-2d}}\chi_{i,\pm}  \\
\pm {iw}{ {\rm e}^{ \k (2\gg -1)+2d}}\chi_{e,\pm} & 1
\end{matrix}\ri] 
e^{-(\varkappa \gg(z)+d(z))\sigma_3} 
\ee 
has the exact same jumps as $\Gamma(z;\l)$ on  $I_i\cup I_e$.

In terms of the Riemann Theta functions $\Th$, the explicit solution to the RHP \ref{modelRHP} is given by
\bea\label{Psi}
\Psi(z;\k)= C_0
\le[\begin{array}{cc}
\ds  \frac { \Theta( \mathfrak u(z)- \mathfrak u(\infty)  - W_0 + W) r(z)}
{\Theta(W-W_0) \Theta(\mathfrak u(z) - \mathfrak u(\infty) - 
W_0)} & 
 \ds \frac { \Theta( -\mathfrak u(z)- \mathfrak u(\infty)  - W_0+ W) r(z)}
 {\Theta(W-W_0)\Theta(-\mathfrak u(z) - \mathfrak u(\infty)  -
W_0)}\cr
\ds  \frac {-\Theta( \mathfrak u(z)+ \mathfrak u(\infty)  -W_0 + W) r(z)}
{\Theta(W-W_0)\Theta(\mathfrak u(z) + \mathfrak u(\infty)  -
W_0)} & 
 \ds \frac {-\Theta( -\mathfrak u(z) + \mathfrak u(\infty)  -W_0+ W) r(z)}
 {\Theta(W-W_0)\Theta(-\mathfrak u(z) +\mathfrak u(\infty) -  W_0)}
 \end{array}\ri],
 \eea
where vectors $W=W(\k)$ and $W_0$ are defined in \eqref{some_not} and $C_0=[\mathbb A^{-1} \nabla \Theta(W_0)]_g$ is the 
last entry of the vector $\mathbb A^{-1} \nabla \Theta(W_0)$. The constant $C_0\neq 0$. 

\bl\label{lem-facts-need}
The endpoints $a_n$, $n\in J$, and infinity (on one of the sheets) are the only zeroes (in $z$) of the
functions  $\Theta\le(-(-1)^k\mathfrak u(z)+(-1)^j\mathfrak u(\infty)+W_0\ri)$, $j,k=1,2$. All these zeroes are simple.
\el

According to Lemma \ref{lem-facts-need}, $\Psi(z;\k)$ is well defined
if  the denominator $\Theta(W-W_0)\neq 0$.

\bth\label{theo-exist}[Thm. 5.3, \cite{BKT2}]
The RHP \ref{modelRHP} has a solution if and only if  $\Theta(W-W_0)\neq 0$.
\et
As a consequence, we obtained the condition \eqref{kappa_n_cond} for the logarithms of approximate eigenvalues.
According to \eqref{round-phi_n}, in order to approximate singular functions, we need to calculate the residues of  
\eqref{Psi}. 

\bp[Symmetry]
\label{symmetry}
If $\Psi(z;\k)$ satisfies the  RHP \ref{modelRHP} then $\det \Psi\equiv 1$ and $\wt \Psi(z)\equiv \Psi(z)$,
where $\wt \Psi(z;\k)=\overline{\Psi(\bar z; \ov \k)}$.
In particular, for $\k\in \R$,  $\Psi_{j1+}(z;\k)= \ov \Psi_{j1-}(z;\k)$ for any $z\in I = I_i\cup I_e$.
\ep

Further analysis of singular functions requires some information about zeroes of the Theta function, given in Section \ref{sec-res-main}

\subsection{Theta divisors and some related results from \cite{BKT2} }\label{sec-res-main}


\bd
\label{prop-K}
Let $a_1$ be a base-point of the Abel map $\mathfrak u(z)$  (see \eqref{Abelmap})  on the hyperelliptic Riemann surface $\Rscr$ 
of $\sqrt{\prod_{j=1}^{2g+2} (z-a_j)}$. Then the vector of Riemann constants $\mathcal K$ is 
\be
\mathcal K = \sum_{j=1}^g \mathfrak u(a_{2j+1}).
\ee
\ed
\bth[\cite{FarkasKra}, p. 308]
\label{generalTheta}
Let ${\bf  f}\in \C^g$ be arbitrary, and denote by $\mathfrak u(p)$ the Abel map extended to the whole Riemann surface.
The (multi-valued) function $\Theta(\mathfrak u(z) - {\bf  f})$ on the Riemann surface either vanishes identically or 
vanishes at $g$ points ${p}_1,\dots, {p}_g$ (counted with multiplicity).
In the latter case we have 
\be
{\bf  f} =\sum_{j=1}^{g} \mathfrak u(p_j) + \mathcal K.
\ee
\et

\br
Description of the vectors ${\bf f}$  that lead to identically vanishing $\Theta(\mathfrak u(z) - {\bf  f})$
is more involved and will not be discussed here.
\er

Let us denote by $\Lambda_\tau = \Z^g + \tau \Z^g\subset \C^g$ the {\em lattice of periods}. 
The {\bf Jacobian} is the quotient $\mathbb J_\tau = \C^g\mod \Lambda_\tau$ and it is a compact torus of real 
dimension $2g$ on account of Theorem \ref{Riemann1}.


\bd\label{def-thetadiv}
The {\bf theta divisor} is the locus ${\bf  e}\in\mathbb J_\tau$ such that
$\Theta({\bf e})=0$. It will be denoted by the symbol $(\Theta)$.
\ed

\bp[Prop.7.1, \cite{BKT2}]
\label{ThetaDiveig}
If $W\in \R^g$ and $W_0$ is given as in \eqref{wukappa}, then
\be
\Theta\le ( W - W_0\ri)=0~~\Longleftrightarrow~~~  W = \sum_{\ell=1}^{g-1}\le( \mathfrak u (p_{{\ell+1}}) -\mathfrak u(a_{j_\ell}) \ri)\mod\ \Z^g ,
\ee 
where  $p_{\ell+1}= (z_{\ell+1}, R_{\ell+1}), \ell =1,\dots, g-1,$ are arbitrary points with $z_{\ell+1}\in [a_{2\ell}, a_{2\ell+1}],$ $\ell =1,\dots, g-2,$ and $z_{g}\in \R\setminus [a_1,a_{2g+2}]$ (i.e. belonging to the cycles $A_{1+\ell}, \ell =1,\dots, g-1$), and $j_\ell \in J = \{1, 5,7, 9,11, \dots,2g-1\}$. 
\ep

\br\label{rem-f_n}
Proposition \ref{ThetaDiveig} explicitly parametrizes the hypersurface $\Theta\le ( W - W_0\ri)=0,\ W\in \R^g$ in terms of $g-1$ points $p_2,\dots, p_g$ belonging to the 
cycles $A_2,\dots, A_g$.
For the special values $\k=\tk_n$, when  the line $W(\k)$ (given by \eqref{wukappa})   intersects with this hypersurface,
we shall denote the correponding points on the cycles $A_2,\dots A_g$ by  $p_2^{(n)}, \dots, p_{g}^{(n)}$ 
with $\vec p_{n}=(p_2^{(n)}, \dots, p_{g}^{(n)})$.
According to \eqref{some_not} and
Theorem \ref{generalTheta},   $\mathbf f_n=  \sum_{j=2}^{g}\mathfrak u(p^{(n)}_j) + \mathcal K$.
For this reason it makes sense to consider ${\bf f}(\vec p) :=   \sum_{j=2}^{g}\mathfrak u(p_j) + \mathcal K$, where $\vec p=(p_2,\dots,p_{g})$, as a function on the (universal cover) of the torus $A_2\times \dots \times A_g$ 
Then we have 
  ${\bf f}_n = {\bf f} (\vec p_n)$.
\er

\bl[Lem. 7.14, \cite{BKT2}]
\label{lemmarespsi}
{\bf (1)} For $\Psi(z;\k)$ from \eqref{Psi} we have
\be\label{res-Psi}
\res{\k=\tk_n} \Psi(z;\k)  = C_0 \le[
\begin{matrix}
\ds \frac{i\pi \Theta\le(\mathfrak u(z) -\mathfrak u(\infty) + {\bf f}(\vec p_n) \ri) r(z)}{\vec \tau_1 \!\!\cdot\!\!\nabla \Theta(
{\bf f}(\vec p_n)) \Theta\le(\mathfrak u(z) -\mathfrak u(\infty)+W_0\ri)} 
&
\ds \frac{i\pi \Theta\le(-\mathfrak u(z) -\mathfrak u(\infty) +{\bf f}(\vec p_n) \ri) r(z)}{\vec \tau_1 \!\!\cdot\!\!\nabla \Theta(
{\bf f}(\vec p_n)) \Theta\le(-\mathfrak u(z) -\mathfrak u(\infty)+W_0\ri)} \\[5ex]
\ds \frac{-i\pi \Theta\le(\mathfrak u(z) +\mathfrak u(\infty) +{\bf f}(\vec p_n)\ri) r(z)}{\vec \tau_1 \!\!\cdot\!\!\nabla \Theta(
{\bf f}(\vec p_n)) \Theta\le(\mathfrak u(z) +\mathfrak u(\infty)+W_0\ri)} 
&
\ds \frac{-i\pi \Theta\le(-\mathfrak u(z) +\mathfrak u(\infty) + {\bf f}(\vec p_n) \ri) r(z)}{\vec \tau_1 \!\!\cdot\!\!\nabla \Theta(
{\bf f}(\vec p_n)) \Theta\le(-\mathfrak u(z) +\mathfrak u(\infty)+W_0\ri)} 
\end{matrix}
\ri].
\ee
${\bf (2)}$  For any $\vec p_n \in A_2\times \dots\times A_g$ the matrix in \eqref{res-Psi} is not identically zero. {\bf(3)} The two rows of the matrix in \eqref{res-Psi} are proportional to each other for any $\vec p_n \in A_2\times \dots\times A_g$.
\el

\bl
\label{technical}
{\bf(1)} The following identities hold for $j=1,2$:
\be
\label{psipsi}
  N_{j}(\vec p_n):=- \frac i{\pi^2}\oint_{B_1}\res{\k=\tk_n} \!\!\! \Psi_{j1}(z;\k)\!\!\! \res{\k=\tk_n} \!\!\! \Psi_{j2}(z;\k)d z
=
\frac {\Theta({\mathbf f}(\vec p_n)+(-1)^j2\mathfrak u(\infty))} {\Theta(W_0+(-1)^j 2\mathfrak u(\infty) )}\frac{ [\mathbb A^{-1} \nabla \Theta(W_0)]_g }
{i \vec \tau_1\!\!\cdot \!\! \nabla \Theta({\mathbf f}(\vec p_n)}.
\ee
{\bf (2)} The function $N_j(\vec p)$ is a (real) analytic function of $\vec p\in A_2\times \dots \times A_g$. It vanishes to second order at $p_{g-1} =\infty_{l}$, where $\infty_l$ is the point at $z=\infty$ on the sheet $l=1,2$, and has no other zeroes.
\el

According to \eqref{round-phi_n}, \eqref{svd-K}, the normalized singular function $\wh f_n(z)$ is propotional to 
\be\label{varps_jn}
\varphi_{n,j}(z)=
{i} \sqrt{w(z)}\res{\l=\l_n} \Gamma_{j2}(z;\l) \frac {1}{\l}, ~~~j=1,2, 
\ee
where at least one of  the latter expressions is not zero. Note that $\varphi_{n,j}$ corresponds to the second term 
of $\phi_{n,j}$ from \eqref{round-phi_n}.

 \bp
 \label{asymptnorms}
The norms in $L^2({I}) $ of the singular functions $\phi_{n,j}$ are given by
 \be
 \label{748}
\|\phi_{n,j}\|^2 =     2{\rm e}^{(-1)^{j+1}2(d_\infty + \tk_{n}\gg_{\infty}) - \tk_n} \le(\pi^2 N_j(\vec p_n)
 + \mathcal O(\tk_n^{-1})\ri),~~~j=1,2.
 \ee
Moreover, $\|\varphi_{n,j}\|^2 =\hf \|\phi_{n,j}\|^2$, where  $\|\varphi_{n,j}\|$ is the $L^2({I_i}) $ norm of
$\varphi_{n,j}$.
 \ep

\bc
\label{doublecover}
 {\bf (1)}
The functions $N_j(\vec p)$ have constant sign on the torus $A_2\times \dots A_g$.
The function $\sqrt{N_j(\vec p)}$ can be defined analytically on the double cover of  $A_2\times \dots A_g$.
{\bf (2)} There exists $\nu>0$ such that for all $ \vec p \in A_2\times \dots A_g$
\be\label{748a}
\max_{j=1,2}|N_j(\vec p)| >\nu.
\ee
\ec


\bc\label{cor-Ups12}[Cor. 7.20, \cite{BKT2}]
Functions 
\be\label{Upsns}
\Upsilon_{j,k}(z;\vec p)=\sqrt{ \frac {\Theta(W_0\!+\!(-1)^j2\mathfrak u(\infty) )}{\Theta(\mathbf f(\vec p)\!+\! (-1)^j2\mathfrak u(\infty))}\times
 \frac {[\mathbb A^{-1} \nabla \Theta(W_0)]_g }{i\vec \tau_1\!\!\cdot \!\! \nabla \Theta(\mathbf f(\vec p)) }}
\frac{\Theta\le((-1)^{k+1}\mathfrak u(z)\! +\!\!(-1)^j\!\mathfrak u(\infty) + \mathbf f(\vec p) \ri) r(z)}
{ \Theta\le((-1)^{k+1}\mathfrak u(z) \!+\!(-1)^j \mathfrak u(\infty)+W_0\ri)},
\ee
$j,k=1,2$, are analytic in $z$ on $Z_0$ and in $\vec p$ on the double covering of the torus  $ A_2 \times \dots A_g$,
where $Z_0 = \bar\C \setminus [a_1,a_{2g+2}]$  together with the boundary points on both sides of the interval $ (a_1,a_{2g+2})$.
Moreover, $\Upsilon_{1,k}(z;\vec p)$ coincides with $\Upsilon_{2,k}(z;\vec p)$, $k=1,2$, on
$Z_0 \times A_2 \times \dots A_g$ modulo factor $(-1)$. 
\ec

\br\label{rem-2Ups}
Note that $(\Upsilon_{j,1})_+(z;\vec p_n)=   \Upsilon^{(j)}(z;\mathbf f_n ) $, the latter  defined in \eqref{Ups}, where $z\in I$.
The subscipt ``$+$''  indicates  that the limiting value on the upper side of $z\in I$ in $Z_0$ is taken.
In view of Corollary \ref{cor-Ups12}, we denote by $\Upsilon_k(z;\vec p)$ a function on $Z_0 \times A_2 \times \dots A_g$
that coincides (modulo sign) with both  $\Upsilon_{1,k}(z;\vec p)$ and $\Upsilon_{2,k}(z;\vec p)$, $k=1,2$. 
Then for each $n\in\N$ we have $(\Upsilon_1)_+(z;\vec p_n) = \Upsilon(z;\mathbf f_n ) $ on $z\in I$, see
Theorem \ref{cor-first}, modulo factor $(-1)$.
\er


\section{Proofs of the technical  lemmas}
\label{proofstechn}
In this section, we use $\wh f_n$ to denote the $n$-th normalized singular function for the system \eqref{svd-def2}, as well
as its analytic continuation on $\C\setminus I_e$. 
It follows from \eqref{svd-def2} that each $\wh f_n$ is purely imaginary on $I_i$ and  defined uniquely modulo the factor $-1$. 
According to \eqref{gammaexact},
\be
\label{hatf_prop}
 \varphi_{n,j}(z) = i \sqrt{w(z)} {\rm e}^{ \tk_n (\gg(z) -(-1)^j \gg_\infty) + (d(z)-(-1)^jd_\infty)} \bigg(\res{\k=\tk_n} \Psi_{j,2}(z;\k) + \mathcal O(\wt \k_n^{-1})\bigg)\
\ee
uniformly on any compact set not intersecting the lenses $\mathcal L_{i}^{(\pm)},~ \mathcal L_e^{(\pm)}$, see Figure \ref{Lenses}, and 
\bea
\label{hatf_proplens}
&\& \varphi_{n,j}(z) = i\sqrt{w(z)}  m_{n,j}
\le[ \res{\k=\tk_n }  
\le(\frac {\pm 1}{i w(z)} \Psi_{j1}(z) {\rm e}^{-\k  (\gg(z) +1) -d(z)}   + \Psi_{j2} (z){\rm e}^{\k \gg(z) + d(z)}
\ri)  + \mathcal O(\wt \k_n^{-1})\ri],\cr
&&\hspace{1cm}{\rm where}~~~ m_{n,j}:=   {\rm e}^{   -(-1)^{j}  \tk_n  \gg(\infty) -(-1)^j d(\infty) },
\eea
uniformly on any compact subset of $\mathcal L_i^{(\pm)}$ 
not containing the endpoints.

We now  define the approximations $\varphi_{n,j}^{(\infty)}(z)$ of $ \varphi_{n,j}(z)$ as 
$\varphi^{(\infty)}_{n,j}=
{i} \sqrt{w(z)}\res{\l=\l_n} \Gamma^{(\infty)}_{j2}(z;\l) \frac {1}{\l}$, $j=1,2$.
Then, according to  \eqref{gammainfty}, 
\bea
\label{hatf_propinfty}
&\& \varphi_{n,j}^{(\infty)}(z) = i \sqrt{w(z)} {\rm e}^{ \tk_n (\gg(z) -(-1)^j \gg_\infty) + (d(z)-(-1)^jd_\infty)} \res{\k=\tk_n} \Psi_{j,2}(z;\k) 
\eea
for $z\in \C \setminus \bigcup_{\pm} \mathcal L_{i}^{(\pm)} \cup \mathcal L_e^{(\pm)}$  and
\bea
\label{hatf_proplensinfty}
&\& \varphi_{n,j}^{(\infty)} (z) = i m_{n,j} \sqrt{w(z)}\res{\k=\tk_n }  
\le(\frac {\pm 1}{i w(z)} \Psi_{j1}(z) {\rm e}^{-\k  (\gg(z) +1) -d(z)}   + \Psi_{j2} (z){\rm e}^{\k \gg(z) + d(z)}
\ri) 
\eea
for $z\in \mathcal L^{(\pm)}_i$ (we will not need an expression in $\mathcal L^{(\pm)}_e$).

\br
It follows from  \eqref{geqm}, \eqref{propertyDelta} and  that Schwarz symmetry of $\gg(z), d(z)$ that
 $\gg_+ + 1  = \frac 1 2  + \frac 1 2 (\gg_+ - \gg_-) = \frac 1 2 + i \Im \gg_+$ on $I_i$ and 
$d_+  = \frac {d_+- d_- - \ln w}2  = i\Im d_+ - \frac 1 2 \ln w$ on $I$. 
Then, taking the $+$ boundary value of \eqref{hatf_proplensinfty} and using
 Proposiotion \ref{symmetry}, we obtain
 the following chain of equalities valid for $z\in I_i$ (we omit the dependence on $z$ for brevity) 
\bea
\varphi_{n,j}^{(\infty)}(z)= i m_{n,j}\sqrt{w} \res{\k=\tk_n }  
\le(\frac {1}{i w} \Psi_{j1+} {\rm e}^{-\k  (\gg_+ +1) -d_+}   + \Psi_{j2+} {\rm e}^{\k \gg_+ + d_+}
\ri)  =\cr
=   m_{n,j} \res{\k=\tk_n }  
\le( \Psi_{j1+} {\rm e}^{-i \k\Im \gg_+  - \frac \k 2 -i\Im d_+}   + i
\ov{ \Psi_{j2-}} {\rm e}^{i\k\Im \gg_+  - \frac \k 2  +i\Im  d_+}
\ri) =\cr
=  m_{n,j}{\rm e}^{-\frac {\tk_n} 2} \res{\k=\tk_n }  
\le(\Psi_{j1+} {\rm e}^{-i \k\Im \gg_+  -i\Im d_+}   -\ov{ \Psi_{j1+}} {\rm e}^{i\k\Im \gg_+    +i\Im  d_+}
\ri) =
\cr
= 2im_{n,j}{\rm e}^{-\frac {\tk_n} 2}  
\Im \le(\res{\k=\tk_n }\Psi_{j1+}   {\rm e}^{-i \tk_n\Im \gg_+  -i\Im d_+} 
\ri). 
\label{imphinfty}
\eea
\er

\bl
\label{lemmab1}
The
functions 
$\prod_{j=1}^{2g+2} (z-a_j)^\frac 1 4\Upsilon_{jk}(z;\vec p)$ are uniformly bounded on the compact
$\bar Z_0\times A_2 \times \dots A_g$.
\el
\begin{proof}
The closure  $\bar Z_0$ of $Z_0$  include the endpoints $a_j$, $j=1,\dots,2g+2$. 
Lemma \ref{technical} and Corollary \ref{cor-Ups12} imply
\be\label{ind_Ups}
\Upsilon_{jk} (z;\vec p)=
\frac{(-1)^j}{\sqrt{N_{j}(\vec p)}}\times
\frac{\Theta\le((-1)^{k+1}\mathfrak u(z)\! +\!\!(-1)^j\!\mathfrak u(\infty) + \mathbf f(\vec p) \ri) r(z)}
{ \Theta\le((-1)^{k+1}\mathfrak u(z) \!+\!(-1)^j \mathfrak u(\infty)+W_0\ri)}.
\ee
Moreover, according to  Corollary \ref{cor-Ups12} and \eqref{748a}, we can always assume that for a given $\vec p
 \in A_2 \times \dots A_g$  we have $\le|\frac{1}{\sqrt{N_{j}(\vec p)}}\ri| < \nu^{-\hf}$. Thus, it remains to 
estimate the second factor in \eqref{ind_Ups}. 

The numerator $\Theta( (-1)^{k+1} \u(z) +\!\!(-1)^j \u(\infty)  + \mathbf f(\vec p))$ is
 analytic (in all variables) on the compact set $(z;\vec p)\in  \bar Z_0 \times A_2 \times \dots A_g$  and, thus, bounded there.
The Theta function in the denominator depends only on $z$. According to Lemma \ref{lem-facts-need},  it 
 vanishes only at infinity (like $z^{-1}$) and at  the $g-1$ points $z = a_j, j\in J$, where it vanishes like $\sqrt{z-a_j}$. 
Taking into account \eqref{spinorh}, we see that the ratio 
$\frac{ r(z)}{ \Theta\le((-1)^{k+1}\mathfrak u(z) \!+\!(-1)^j \mathfrak u(\infty)+W_0\ri)}$ is bounded
on $\bar Z_0$ away from the endpoints $a_j$, and behaves like  $O(z-a_j)^{-\frac 14}$ near each endpoint $a_j$, $j=1,\dots, 2g+2$.
Thus, the statement of the lemma is proven.
\end{proof}

\br\label{rem-Ups-phi} 
According to Lemmas \ref{technical},\ref{lemmarespsi}, 
\be\label{resPsiUps}
\Upsilon_{j,k}(z;\vec p_n)=\frac{\res{\k=\tk_n }\Psi_{jk}(z;\k)}{\pi\sqrt{N_j(\vec p_n)}},~~~j,k=1,2.
\ee
Thus, \eqref{imphinfty} implies 
that $\varphi_{n,j}^{(\infty)}$ belongs to $L^2(I_i)$. 
\er
Let $ \Jscr^\o$ denote the $\o$ neighborhood of 
{ the endpoints of $I$.}

\bl\label{lem-est-gaps}
 For any $\o>0$ there exists some $c_\o>0$ such that 
\be\label{est_whf_n}
|\wh f_n(z)|\leq \frac {c_\o}{
{1 + |z|^\frac 1 2}}{\rm e}^{ \k_n (\Re\gg(z)+ \frac 1 2 )}~~~{\rm on}~~~\bar\C\setminus\Jscr^\o.
\ee
\el
\begin{proof}
As it was mentined in Section \ref{sec-res-main}, $\wh f_n=  \varphi_{n,j}/\|\varphi_{n,j}\|$, $j=1,2$, provided  
$\|\varphi_{n,j}\|>0$. Note that for every $n\in\N$  at least one of $\|\varphi_{n,j}\|>0$. Then,
{ using  the estimate \eqref{err_est} for $\mathcal E(z;\k)$  and taking the residue of \eqref{gammaexact}, we have
\be
\label{gap-approx}
\wh f_n(z)= 
  i{\rm e}^{ \tk_n (\gg(z)+ \frac 1 2 ) +d(z)} \le(\sqrt{ w(z)} \Upsilon_{2}(z;\vec p_n)   +\frac{\sqrt{|w(z)|} \mathcal O(\tk_n^{-1})}{1 + |z|}\ri)
\ee
uniformly on closed 
subsets of $\C \setminus \Jscr^\o$. Here we also used Remarks \ref{rem-Ups-phi} and 
\ref{rem-2Ups}.
Near $z=\infty$ the function $\Upsilon_{2}(z;\vec p_n)$ has behavior 
\be\label{est-Upsi}
\Upsilon_{2}(z;\vec p_n) = \frac {K(\vec p_n)}{z} + \mathcal O(z^{-2}), 
\ee
see RHP \ref{modelRHP} and \eqref{resPsiUps}, with some constant $K(\vec p_n)>0$.
Since $K(\vec p_n)$ is continuous on the compact set  $\vec p_n\in A_2\times \dots \times A_g$,  we 
conclude that there is $\wh K>0$ and a neighborhood of $z=\infty$ such  that  $|\Upsilon_{2}(z;\vec p_n) |\leq \frac {\wh K}{|z|}$ in this neighborhood 
for
all $n\in\N$. Then, according to Lemma \ref{lemmab1}, there exists some $K>0$, such that
\be
|\Upsilon_{2}(z;\vec p_n)| \leq \frac {K}{1+|z|}
\ee
on  $\C \setminus \Jscr^\o$ 
 uniformly in $n\in\N$. The statement thus follows from \eqref{gap-approx}.}
\end{proof}

\br\label{rem-subtlety}
The subtlety in proving the accuracy in \eqref{gap-approx} is that $\wh f_n$ is obtained by dividing  \eqref{hatf_propinfty} by  \eqref{748} and, although
$\max_{j=1,2} |N_{j}(\vec p_n)|$ is separated from zero, each of the sequences $|N_{1}(\vec p_n)|$,  $|N_{2}(\vec p_n)|$ 
is, in general, not.
However, for each $n\in\N$ we could always use the particular choice of $j$ that provides the said maximum, which guarantees the uniformity of the estimate. 
\er



The following corollary is a  direct consequence of   Corollary \ref{cor-Ups12}.

\bc\label{cor-bound-gaps}
For any $\o>0$ and any closed interval $\Iscr \subset \R\setminus\Jscr^\o$, the functions 
\be\label{munu}
\mu_\Iscr(\vec p)=\max_{z\in\Iscr} |\Upsilon_{2}(z;\vec p)|,~~~\nu_\Iscr(\vec p)=\max_{z\in\Iscr} |\frac{\partial}{\partial z}\Upsilon_{2}(z;\vec p)|,
\ee
are continuous  on  $ A_2 \times \dots \times A_g$.
\ec

{\bf Proof of Lemma \ref{lem-seq-intervals}.} 
Let us choose $\o>0$ so that ${\bf J}\setminus \Jscr^\o$ contains some segment $\Iscr$. We construct intervals $J_n\subset \Iscr$, $n\in\N$.
In view of 
\eqref{gap-approx}
and \eqref{acc_sing_val}, it is sufficient to construct $ J_n$ so that $|\Upsilon_2(z;\mathbf f_n) |$ instead
of $|f_n e^{-\k_n (\gg(z)+\hf)}|$ 
will be  separated from zero  on $J_n$ (uniformly in $n$).
Let
\be\label{min-max}
\mu_*=\min_{\vec p\in A_2 \times \dots \times A_g}\mu_\Iscr(\vec p),~~~\nu_*=\max_{\vec p\in A_2 \times \dots \times A_g}\nu_\Iscr(\vec p),
\ee
and let the maximum $\mu_\Iscr(\vec p)$ of $| \Upsilon_2(z;\vec p)|$ in $z\in\Iscr$  be attained at some  $z_{\vec p}\in\Iscr$.
Obviously, $\mu_*>0$ and $\nu_*<\infty$.
Let us now define the intervals $J_n$
by
\be\label{J_n}
J_n= \le(z_{\vec  p^{(n)}}-\frac{\mu_*}{2\nu_*}, z_{\vec  p^{(n)}}+\frac{\mu_*}{2\nu_*}\ri)\cap \Iscr.
\ee
Then the length of each $J_n$ is at least $\min(\frac{\mu_*}{2\nu_*},|\Iscr|)$, where $|\Iscr|$ is the length of $\Iscr$. Then, according to \eqref{munu}, \eqref{min-max},
\be\label{lower_bound} 
|\Upsilon_2(z;\vec p^{(n)})|\geq \frac{\mu_*}{2}
\ee
for all $z\in J_n$. 
Thus, we completed the proof Lemma \ref{lem-seq-intervals}.

{\bf Proof of Lemma \ref{lem-norm-tf}.} 
The norm of $ \varphi_{n,j}^{(\infty)}$ in $L^2(I_i)$ (here $\wt \l_n = {\rm e}^{-\tk_n }$ is the approximate singular-value) 
{is given by }
\be
\int_{I_i}\le(\res{\l=\wt\l_n}  \Gamma^{(\infty)}_{j2}(z;\l) \frac {1}\l\ri)^2 \!\!\!\! {w(z)} dz=-i  \int_{I_i} \le(\res{\l=\wt \l_n} \l J(\Gamma^{(\infty)}_{j1}(z;\l))\frac {1}\l\ri)  \le(\res{\l=\wt \l_n}  \Gamma^{(\infty)}_{j2}(z;\l) \frac {1}\l\ri) dz,
\label{c11}
\ee
where $J(F) = F_+-F_-$.
We can thus deform the two contributions from the $\pm $ boundary values to $\pa \L^{(\pm)}_i$ which consists of arcs joining the consecutive endpoints of $I_i$  (in the formula below, we omit the reference to the dependence on $\k, z$ for brevity): 
\bea
&\&
-i  \int_{I_i} \le(\res{\l=\wt \l_n} \l J(\Gamma^{(\infty)}_{j1}(z;\l))\frac {1}\l\ri)  \le(\res{\l=\wt \l_n}  \Gamma^{(\infty)}_{j2}(z;\l) \frac {1}\l\ri) dz
=\cr
=&\&-i m_{n,j}^2 {{\rm e}^{- {\tk_n}}}\sum_{\pm} (\pm)\int_{ \pa \L_i^{(\pm)} }
\res{\k=\tk_n } \Psi_{j1}{\rm e}^{-\k \gg - d}  \res{\k=\tk_n }  
\le(\frac {\pm 1}{i w} \Psi_{j1} {\rm e}^{-\k  (\gg +1) -d}   + \Psi_{j2} {\rm e}^{\k \gg + d}
\ri) dz = \nonumber\\
&\& =-i m_{n,j}^2{{\rm e}^{- {\tk_n}}}
\bigg(\oint_{B_1}  \res{\k=\tk_n } \Psi_{j1}(z;\k) \res{\k=\tk_n }   \Psi_{j2}(z;\k) dz + 
\label{A15}
\\&\& \hspace{1cm}+
\sum_{\pm} \int_{\pa \mathcal L_i^{(\pm)} } \le(\res{\k=\tk_n } \Psi_{j1}(z;\k) \ri)^2
\frac{{ \rm e}^{-\tk_n   (2\gg(z) +1) -2d(z)} dz}{iw(z)}\bigg). \label{A16}
\eea
The expression \eqref{A15} is precisely $m_{n,j}^2N_{n,j}$ from \eqref{psipsi}.
The remaining  terms on line \eqref{A16} contribute to order $\mathcal O(\tk_n ^{-1})$ as we now explain. Indeed,
according to \eqref{resPsiUps} and Lemma \ref{lemmab1},  there exists some $C>0$, such that 
the integrals in \eqref{A16} are bounded   by 
\be
\int_{\partial L_i^{(\pm)} } \frac {C}{|\sqrt{\prod_{j=1}^{2g+2} (z-a_j)}|}\frac{{\rm e}^{-\tk_n \Re (2\gg(z) +1) -2\Re d(z)} }{|w(z)|} \,|dz|
\ee
uniformly in $n\in\N$.  Using $\Re (2\gg(z) +1)  = C_j|z-a_j|^{\frac 1 2 }(  1 + \mathcal O(|z-a_j|))$, see \eqref{some_not},
we can estimate \eqref{A16}
 to be of order $\mathcal O(\k^{-1})$ as $\k\to  + \infty$.
Thus, according to Lemma \ref{technical}, 
we have proved 
\be
\label{D5}
\le\|  \varphi_{n,j}^{(\infty)} \ri\|^2_{I_i} = m_{n,j}^2  {{\rm e}^{- {\tk_n}}}\pi^2N_{j}(\vec p_n) \le(1  
 + \mathcal O(\tk_n^{-1})\ri ).
 \ee


Using \eqref{tfn}, \eqref{imphinfty} and \eqref{resPsiUps}, we obtain
\be\label{hatfn-exp}
\wt f_n(z) =\frac {\varphi_{n,j}^{(\infty)}(z)}{m_{n,j} {{\rm e}^{-\frac 1 2 {\tk_n}}}\pi\sqrt{N_{j}(\vec p_n)}},
\ee
where the right hand side  does not depend on $j=1,2$.
Now \eqref{D5}, \eqref{hatfn-exp} and \eqref{in1} imply $\|\tf_n(z)\|_{I_i}=1+ \mathcal O(n^{-1})$.

\bibliographystyle{amsalpha}
\bibliography{bibexport}


\end{document}

%% file: Homology2.pdf_t
\begin{picture}(0,0)%
\includegraphics{Homology2.pdf}%
\end{picture}%
\setlength{\unitlength}{3947sp}%
\begingroup\makeatletter\ifx\SetFigFont\undefined%
\gdef\SetFigFont#1#2#3#4#5{%
  \reset@font\fontsize{#1}{#2pt}%
  \fontfamily{#3}\fontseries{#4}\fontshape{#5}%
  \selectfont}%
\fi\endgroup%
\begin{picture}(18494,5369)(54,-7433)
\put(6151,-7111){\makebox(0,0)[lb]{\smash{{\SetFigFont{14}{16.8}{\familydefault}{\mddefault}{\updefault}{\color[rgb]{0,0,0}$a_4$}%
}}}}
\put(1651,-7111){\makebox(0,0)[lb]{\smash{{\SetFigFont{14}{16.8}{\familydefault}{\mddefault}{\updefault}{\color[rgb]{0,0,0}$a_1$}%
}}}}
\put(3376,-7111){\makebox(0,0)[lb]{\smash{{\SetFigFont{14}{16.8}{\familydefault}{\mddefault}{\updefault}{\color[rgb]{0,0,0}$a_2$}%
}}}}
\put(4651,-7111){\makebox(0,0)[lb]{\smash{{\SetFigFont{14}{16.8}{\familydefault}{\mddefault}{\updefault}{\color[rgb]{0,0,0}$a_3$}%
}}}}
\put(3826,-3736){\makebox(0,0)[lb]{\smash{{\SetFigFont{20}{24.0}{\familydefault}{\mddefault}{\updefault}{\color[rgb]{0,0,0}$A_1$}%
}}}}
\put(6676,-3736){\makebox(0,0)[lb]{\smash{{\SetFigFont{20}{24.0}{\familydefault}{\mddefault}{\updefault}{\color[rgb]{0,0,0}$A_2$}%
}}}}
\put(5026,-2986){\makebox(0,0)[lb]{\smash{{\SetFigFont{20}{24.0}{\familydefault}{\mddefault}{\updefault}{\color[rgb]{0,0,0}$B_1$}%
}}}}
\put(8251,-3436){\makebox(0,0)[lb]{\smash{{\SetFigFont{20}{24.0}{\familydefault}{\mddefault}{\updefault}{\color[rgb]{0,0,0}$B_2$}%
}}}}
\put(14851,-3511){\makebox(0,0)[lb]{\smash{{\SetFigFont{20}{24.0}{\familydefault}{\mddefault}{\updefault}{\color[rgb]{0,0,0}$B_g$}%
}}}}
\put(16726,-3661){\makebox(0,0)[lb]{\smash{{\SetFigFont{20}{24.0}{\familydefault}{\mddefault}{\updefault}{\color[rgb]{0,0,1}$A_g$}%
}}}}
\put(15751,-7111){\makebox(0,0)[lb]{\smash{{\SetFigFont{14}{16.8}{\familydefault}{\mddefault}{\updefault}{\color[rgb]{0,0,0}$a_{2g+2}$}%
}}}}
\put(14026,-7111){\makebox(0,0)[lb]{\smash{{\SetFigFont{14}{16.8}{\familydefault}{\mddefault}{\updefault}{\color[rgb]{0,0,0}$a_{2g+1}$}%
}}}}
\put(13126,-3811){\makebox(0,0)[lb]{\smash{{\SetFigFont{20}{24.0}{\familydefault}{\mddefault}{\updefault}{\color[rgb]{0,0,0}$A_0$}%
}}}}
\put(2401,-3436){\makebox(0,0)[lb]{\smash{{\SetFigFont{20}{24.0}{\familydefault}{\mddefault}{\updefault}{\color[rgb]{0,0,0}$B_0$}%
}}}}
\put(12526,-7186){\makebox(0,0)[lb]{\smash{{\SetFigFont{14}{16.8}{\familydefault}{\mddefault}{\updefault}{\color[rgb]{0,0,0}$a_{2g}$}%
}}}}
\end{picture}%

%% file: Lenses.pdf_t
\begin{picture}(0,0)%
\includegraphics{Lenses.pdf}%
\end{picture}%
\setlength{\unitlength}{3947sp}%
\begingroup\makeatletter\ifx\SetFigFont\undefined%
\gdef\SetFigFont#1#2#3#4#5{%
  \reset@font\fontsize{#1}{#2pt}%
  \fontfamily{#3}\fontseries{#4}\fontshape{#5}%
  \selectfont}%
\fi\endgroup%
\begin{picture}(15330,4871)(361,-7737)
\put(4651,-4336){\makebox(0,0)[lb]{\smash{{\SetFigFont{14}{16.8}{\familydefault}{\mddefault}{\updefault}{\color[rgb]{0,0,0}$a_{3}$}%
}}}}
\put(601,-4336){\makebox(0,0)[lb]{\smash{{\SetFigFont{14}{16.8}{\familydefault}{\mddefault}{\updefault}{\color[rgb]{0,0,0}$a_{1}$}%
}}}}
\put(7276,-4336){\makebox(0,0)[lb]{\smash{{\SetFigFont{14}{16.8}{\familydefault}{\mddefault}{\updefault}{\color[rgb]{0,0,0}$a_4$}%
}}}}
\put(15151,-4336){\makebox(0,0)[lb]{\smash{{\SetFigFont{14}{16.8}{\familydefault}{\mddefault}{\updefault}{\color[rgb]{0,0,0}$a_{2g+2}$}%
}}}}
\put(3301,-4336){\makebox(0,0)[lb]{\smash{{\SetFigFont{14}{16.8}{\familydefault}{\mddefault}{\updefault}{\color[rgb]{0,0,0}$a_{2}$}%
}}}}
\put(12376,-4336){\makebox(0,0)[lb]{\smash{{\SetFigFont{14}{16.8}{\familydefault}{\mddefault}{\updefault}{\color[rgb]{0,0,0}$a_{2g+1}$}%
}}}}
\put(8251,-4336){\makebox(0,0)[lb]{\smash{{\SetFigFont{14}{16.8}{\familydefault}{\mddefault}{\updefault}{\color[rgb]{0,0,0}$a_{2g-1}$}%
}}}}
\put(11101,-4336){\makebox(0,0)[lb]{\smash{{\SetFigFont{14}{16.8}{\familydefault}{\mddefault}{\updefault}{\color[rgb]{0,0,0}$a_{2g}$}%
}}}}
\put(1651,-4486){\makebox(0,0)[lb]{\smash{{\SetFigFont{20}{24.0}{\familydefault}{\mddefault}{\updefault}{\color[rgb]{0,0,0}$\mathcal L_e^{(-)}$}%
}}}}
\put(1651,-3661){\makebox(0,0)[lb]{\smash{{\SetFigFont{20}{24.0}{\familydefault}{\mddefault}{\updefault}{\color[rgb]{0,0,0}$\mathcal L_e^{(+)}$}%
}}}}
\put(5851,-3661){\makebox(0,0)[lb]{\smash{{\SetFigFont{20}{24.0}{\familydefault}{\mddefault}{\updefault}{\color[rgb]{0,0,0}$\mathcal L_i^{(+)}$}%
}}}}
\put(5776,-4486){\makebox(0,0)[lb]{\smash{{\SetFigFont{20}{24.0}{\familydefault}{\mddefault}{\updefault}{\color[rgb]{0,0,0}$\mathcal L_i^{(-)}$}%
}}}}
\put(9526,-3661){\makebox(0,0)[lb]{\smash{{\SetFigFont{20}{24.0}{\familydefault}{\mddefault}{\updefault}{\color[rgb]{0,0,0}$\mathcal L_i^{(+)}$}%
}}}}
\put(13576,-4486){\makebox(0,0)[lb]{\smash{{\SetFigFont{20}{24.0}{\familydefault}{\mddefault}{\updefault}{\color[rgb]{0,0,0}$\mathcal L_e^{(-)}$}%
}}}}
\put(13576,-3661){\makebox(0,0)[lb]{\smash{{\SetFigFont{20}{24.0}{\familydefault}{\mddefault}{\updefault}{\color[rgb]{0,0,0}$\mathcal L_e^{(+)}$}%
}}}}
\put(9526,-4486){\makebox(0,0)[lb]{\smash{{\SetFigFont{20}{24.0}{\familydefault}{\mddefault}{\updefault}{\color[rgb]{0,0,0}$\mathcal L_i^{(-)}$}%
}}}}
\end{picture}%